\documentclass[prb,aps,twocolumn,showpacs]{revtex4-1}
\usepackage{amsmath,amssymb,bm,graphicx,dsfont,color}
\usepackage[latin9]{inputenc}
\usepackage{amsmath}
\usepackage{slashed}
\usepackage{dsfont}
\usepackage{amstext}
\usepackage{amssymb}
\usepackage{amsbsy}
\usepackage{amsthm}
\usepackage{epsfig}
\usepackage{graphicx}
\usepackage{bbm}
\usepackage{hyperref}
\usepackage{color}
\usepackage{array}
\usepackage{cancel}
\usepackage{multirow}
\usepackage{array}

\usepackage[normalem]{ulem}

\PassOptionsToPackage{linktocpage}{hyperref}
\hypersetup{%
    pdfborder = {0 0 0}
}

\newcommand{\<}{\langle}

\renewcommand{\>}{\rangle}
\renewcommand{\(}{\left(}
\renewcommand{\)}{\right)}
\renewcommand{\[}{\left[}
\renewcommand{\]}{\right]}
\renewcommand{\v}[1]{\mathbf{#1}} 

\newcommand{\bs}[1]{\boldsymbol{#1}}
\renewcommand{\d}{\partial}

\newcommand{\Z}{\mathbb{Z}}
\newcommand{\T}{\mathcal{T}}

\begin{document}
 
\title{Protection of topological order by symmetry and many-body localization}
\author{Andrew C. Potter and Ashvin Vishwanath}
\affiliation{Department of Physics, University of California, Berkeley, CA 94720, USA}
 \date{\today}
\begin{abstract}
In closed quantum systems, strong randomness can localize many-body excitations, preventing ergodicity.  An interesting consequence is that high energy excited states can exhibit quantum coherent properties, such as  symmetry protected topological (SPT) order, that otherwise only occur in equilibrium ground states.  Here, we ask: which types of SPT orders can be realized in highly excited states of a many-body (MB) localized system? We argue that this question is equivalent to whether an SPT order can be realized in an exactly solvable lattice model of commuting projectors. This perspective enables a sharp definition of MB localizability. Using this criterion, it is straightforward to establish that whereas all bosonic SPTs in spatial dimensions $d=1,\,2,\,3$ are MB localizable, chiral phases (e.g. quantum Hall fluids) are not.  We also show that  free fermion SPTs in $d >1$  (including topological insulators and superconductors) cannot be localized if interactions are weak. A key question is whether strong interactions can render them MB localizable, which we study in the context of a class of $d=2$ topological superconductors.  Using a decorated domain wall (DDW) approach  we show that some phases in this class are MB localizable, when they correspond to  bosonic SPT orders.  However, a similar DDW approach faces a fatal obstruction to realizing certain intrinsically fermionic SPT orders, an issue we argue may persist beyond this specific construction.  
\end{abstract}
\maketitle


\section{Introduction}
Topological phases often exhibit unconventional edge states with unusual properties - for example the chiral edges of the quantum Hall phases, Majorana end states of topological superconducting wires, and the helical edges of topological insulators.  In equilibrium settings, such topological edge states are protected by a bulk gap at zero temperature, but are, strictly speaking, destroyed at any finite temperature due to thermally activated excitations.  An interesting exception arises in well-isolated and strongly disordered systems in which all excitations are localized, preventing the system from reaching thermal equilibrium - a phenomena known as many-body localization (MBL).\cite{Fleishman1980,Gornyi2005,BAA,PalHuse}  This raises the intriguing possibility of harnessing coherent topological edge states in arbitrarily ``hot" systems without the need for careful cooling.\cite{BauerNayak,Bahri2013,MBLQuantumOrder,AnushyaMBLSPT,RNACP}

The possibility of MBL protection of topological edge states was recently explored in certain one-dimensional topological superconducting wires\cite{MBLQuantumOrder} and symmetry protected phases\cite{Bahri2013}, as well as in the two dimensional symmetry-protected phase with Ising-like $\Z_2$ symmetry\cite{AnushyaMBLSPT} (Levin-Gu model\cite{LevinGu}).  Additionally, MBL was shown to protect intrinsic topological order in toric code type models of discrete gauge theories.\cite{MBLQuantumOrder,BauerNayak} These studies have shown that the excited states of MBL systems have the properties of quantum ground-states, including low entanglement and certain quantum orders. In this paper we would like to ask the converse question: which properties of a quantum ground-state be extended to all excited states in an MBL system? In particular, we aim to understand the scope of possible topological distinctions among different MBL phases, and whether ground-state topology may be protected from thermal decoherence by MBL.


To this end, we develop a more systematic understanding of the localizablity of topological phases, focusing primarily on ``integer" topological- and symmetry protected topological phases (SPTs).  Building on ideas of Refs.~\onlinecite{VoskAltmanDynamicalRG,Serbin2013,BauerNayak}, we use the existence of a complete set of local conserved quantities as the defining feature of an MBL phase, and construct a sharp criterion for whether a phase is many-body localizable (MB localizable). Specifically, we define a phase to be MB localizable if it can be continuously connected (i.e. without encountering a delocalization transition, or equivalently by a finite depth unitary circuit) to one whose entire eigenspectrum is completely characterized by a set of strictly local conserved quantities. Here, we consider only ``full" MBL in which the entire spectrum belongs to the same MBL phase.  In practice, this criterion for MB localizability is equivalent to asking whether a phase may be represented by an exactly solvable, local commuting projector Hamiltonian.  

Apart from the connection to MBL, there are several motivations for constructing such exactly solvable lattice models for topological phases. Firstly, exactly solvable lattice models play an important foundational role in the conceptual development of topological and symmetry protected topological (SPT) phases \cite{AKLT,ToricCode,Kitaev,Kitaev2006,StringNet,LevinGu,WalkerWang}, serving as a proof-of-principle for the existence of new topological phases, and elucidating their properties through simple exact calculations. Moreover, the question of MB localizability is tightly connected to questions about which phases of matter can be efficiently represented by a classical computer. Namely, all eigenstates of a system described by a local commuting projector Hamiltonian can be represented as tensor product states (TPS's) with finite bond dimension.\cite{PEPsReview} Conversely, given a TPS description one can essentially always construct an exactly solvable parent Hamiltonian (given a certain ``injectivity" condition\cite{ParentH1,Garcia1,Garcia2}). Lastly, the construction of exactly solvable models may assist in the efficient simulation of interacting SPT edge states. Typically, simulating the $d$ dimensional edge of a $d+1$ dimensional SPT requires simulating the full $d+1$ dimensional system, due to the anomalous properties of the edge, adding substantial computational overhead and essentially prohibiting the direct simulation of surface states of 3D SPTs. However, the construction of solvable models for bosonic SPTs revealed the possibility of representing the $d$ dimensional edge of a $d+1$ dimensional bosonic SPT as a purely $d$-dimensional lattice model, just with a slightly unconventional symmetry action. If possible, analogous constructions for fermionic SPTs would be highly desirable, permitting for example the efficient simulation of interacting quantum spin Hall edges or 3D topological insulator surface states.

With these motivations in mind and a sharp criterion for MB localizability in hand, we show that chiral phases such as integer and fractional quantum Hall systems are fundamentally not MB localizable.  We then go on to investigate the MB localizability of non-chiral higher dimensional SPT phases.  By suitably modifying the cohomology\cite{CohomologyScience,CohomologyPRB} and Walker-Wang constructions\cite{WalkerWang,Burnell2013} we show that all bosonic SPTs are localizable in all physical dimensions, $d=1,2,3$.   

Next, we turn to the localization of fermionic phases.  Unlike bosonic SPT phases, fermionic SPTs permit a noninteracting description in terms of topological band theory.  However, we establish a fundamental obstruction to localizing such topological bands while preserving symmetry; this generalizes the well known Wannier obstruction for quantum Hall phases\cite{Thouless1984} and the related time-reversal symmetry-based obstruction for topological insulators.\cite{Z2WannierObstruction}  We then argue, that the same obstruction is not necessarily present in interacting fermion systems - and construct a particular example of a localizable \emph{interacting} SPT of fermions using the decorated domain wall (DDW) approach of Ref.~\onlinecite{ChenDDW}.  Interestingly, this DDW construction fails for most fermionic SPTs except for a special set that are topologically equivalent to bosonic SPTs (once interactions are present).  A related obstruction is present for Walker-Wang model based constructions of fermionic SPTs - suggesting the interesting possibility that certain intrinsically fermionic SPTs fundamentally resist localization in dimensions higher than one.

\section{Defining Many-Body Localizability}
Our first step is to formalize the notion of MB localizability.  A defining feature of MBL systems is local integrability.\cite{VoskAltmanDynamicalRG,Serbin2013,BauerNayak}  A system is defined as locally integrable if (almost) all its eigenstates can be labeled by an extensive set of well localized conserved quantities or ``integrals of motion".  Here, well-localized near a position $\v{r}$ implies that the overlap between the integrals of motion and microscopic lattice degrees of freedom decays exponentially with distance from $\v{r}$.  The definition also allows for a measure zero (in the limit of infinite system size) set of ``resonances", states which are not eigenstates of these integrals of motion.\cite{BauerNayak,Imbrie2014} 

Familiar equilibrium phases of matter such as symmetry breaking phases (ferromagnets, superfluids, etc...) are typically defined as the equivalence classes of quantum ground-states that can be continuously deformed into a representative ``RG fixed point" product state without passing through a phase transition.   By analogy, it is natural to define an MBL phase as one whose Hamiltonian can be deformed into a strictly locally integrable Hamiltonian (one with a complete set of conserved quantities that have strictly bounded extent with no exponential tails) without passing through a delocalization transition.\cite{BauerNayak} Here, we make the natural (though not rigorously proven) assumption that any system with exponentially well localized conserved quantities can be smoothly connected (e.g. to arbitrary precision by a finite depth unitary circuit) to a strictly localized one.

As a concrete example, consider a Heisenberg chain in strong random field: $H_\text{XXX}=-\sum_i\(h_i\sigma^z_i+J\bs{\sigma}_i\cdot\bs\sigma_{i+1}\)$. Here, for $\text{var}(h) \gg J^2$, the system is in an MBL phase with eigenstates labeled by emergent integrals of motion\cite{Serbin2013}, $\tau^z_i\approx \sigma_i^z+\dots$, where $(\dots)$ indicates small ``dressing" of $\sigma^z_i$ by other nearby spin operators (with coefficients that decay exponentially with distance from site $i$). It is quite natural to expect these dressed integrals of motion, $\tau_i^z$ can be mapped into the bare physical spins $\sigma_i^z$ to exponential-in-depth accuracy using a finite depth unitary circuit. Hence, we can define the MBL paramagnetic phase of this model by those states that are connected (e.g. to desired accuracy by a finite depth unitary circuit) to the strictly local product eigenstates of $H_\text{XXX}$ with $J=0$, in exact analogy to how ordinary clean, $T=0$ paramagnetic or symmetry broken phase are defined by adiabatic connection to a representative unentangled product state.

Then, MBL phases may be represented as equivalence classes of models that are unitarily equivalent to a representative set of strictly local conserved quantities.  Without loss of generality we may take these conserved quantities as eigenvalues of a complete set of commuting projector operators, $\{\Pi_{\v{r},a}\}$, that have bounded support in the region of position $\v{r}$, and which satisfy $\Pi_{\v{r},a}^2=\Pi_{\v{r},a}$.  From this set of of projectors, we may construct a family of ``fixed point" Hamiltonians for the MBL phase:
\begin{align}
H[\lambda] = \sum_{\v{r},a=1\dots D} \lambda_{\v{r},a}\Pi_{\v{r},a}
\label{eq:HCP}
\end{align}
Here $a=1\dots D$ is a band-index, and to achieve a stable MBL phase $\{\lambda_{\v{r},a}\}$ should be random set of strongly random coefficients chosen such that the variance in of $\lambda_{\v{r},a}$ for nearby $\v{r}$ is much larger than their typical value.  The latter property ensures that the many-body localized character of the system is perturbatively stable to weak breaking of the strictly local integrability.  We define any system that is continuously connected to such an fixed point MBL Hamiltonian as {\bf many-body (MB) localizable}. 

Then, a systematic understanding of MBL phases can be obtained from studying the simpler fixed point Hamiltonians $H[\lambda]$.  In this paper, we are interested in the case where each eigenstate, $|\{n_{\v{r},a}\}\>$,  of $H[\lambda]$ has the properties of a topological or symmetry protected topological ground-state, where 
$\Pi_{\v{r},a}|\{n_{\v{r},a}\}\>=n_{\v{r},a}|\{n_{\v{r},a}\}\>$ and $n_{\v{r},a}\in\{0,1\}$.  For simplicity, we restrict our attention to ``integer" topological states without fractionalized excitations.  These could include ``intrinsic" topological phases, or symmetry protected topological (SPT) phases.  In the latter case, we further demand that the projectors be invariant under the action of the symmetry group $G$ protecting the topological phase. 

Though antiunitary (e.g. time-reversal) and continuous (e.g. charge-conservation) symmetries are more common for electronic systems, these present additional theoretical complications, and for this reason, unless otherwise specified, we will limit our analysis to SPTs protected by discrete, unitary symmetry groups.

Note, also, that our definition allows only for ``full" many-body localization, in which the entire many-body spectrum is labeled by exponentially well localized integrals of motion. This leaves aside more exotic scenarios, such as the possibility of SPT orders that can only arise systems with a a partially MB localized spectrum, separated by thermal eigenstates by a MB mobility edge.

The above arguments show that the question of which topological phases may be protected by MBL is equivalent to the question of which topological phases may be represented by commuting projector Hamiltonians.  Interestingly, these issues are closely connected to quantum information based questions about how to efficiently represent quantum systems on a classical computer.  Specifically, since the ground-state (in fact any eigenstate\cite{Chandran2014}) of a commuting projector Hamiltonian can be represented by a tensor product state (TPS) wave-function with bounded bond dimension, any MB localizable system permits an efficient TPS representation.  The converse is also true so long as the tensors satisfy a certain injectivity condition.\cite{ParentH1,Garcia1,Garcia2}

\section{Preliminaries}
We are now in a position to understand whether or not a large class of topological phases permit MB localization. We begin with some simple examples that are relatively straightforward extensions of previously developed methods, and then turn to the less well developed question of whether fermionic SPT phases may be MB localized.

\subsection{Chiral Phases}
In two dimensions there are a variety of chiral phases that are intrinsically topological (in the sense of not requiring symmetry protection) but not fractionalized (i.e. which have unique ground-state on a torus) - including integer quantum Hall systems and chiral topological superconductors.
However, as pointed out in Ref.~\onlinecite{Kitaev2006,Lin2014}, there is a fundamental obstruction to obtaining a commuting projector Hamiltonian of the form in Eq.~\ref{eq:HCP} for topological phases with chiral edge modes.  Chiral phases are partially characterized by a chiral central charge $c_c$ that determines the imbalance between left and right moving edge modes.  Applying a temperature difference $\Delta T$ across a finite sample of chiral phase produces a thermal Hall current density $j_H = c_c\frac{\pi^2}{3}\frac{k_\text{B}^2}{h}T\Delta T$.  While only the edge carries uniform heat current, this must be accomplished by having circulating thermal currents in the bulk, which cancel to give no uniform contribution except at the edge.  However, the heat current density $j_E(\v{r},n;\v{r}',m)\sim [\Pi_{\v{r}',m},\Pi_{\v{r},n}]=0$  vanishes identically for commuting projector Hamiltonians.  In the present context, this proof demonstrates that chiral phases cannot be MB localized.\cite{Kitaev2006,Lin2014} Furthermore, recent work\cite{RNACP} strongly suggests that the chiral obstruction to localization also inevitably leads to ergodicity and thermalization in interacting chiral phases. 

While chiral edge modes present a fundamental obstacle to MB localization, one might expect that SPT phases, which by definition are smoothly connected (in the absence of symmetry) to trivially localized insualtors and hence have non-chiral edges, may be MB localized. Indeed, we will see that many (though likely not all) SPTs are MB localizable.

\subsection{Bosonic SPTs}
All known SPT phases in $d=1,2,3$, arising from bosonic degrees of freedom are MB localizable. Specifically, group cohomology methods used to classify bosonic SPTs with discrete, unitary symmetries enable an explicit construction of commuting projector Hamiltonians with bosonic SPT grounds states.\cite{CohomologyPRB,CohomologyScience}  As written, these commuting projectors form only an incomplete set, which uniquely specify an SPT ground-state but leave a highly degenerate excited state spectrum.  However, as shown in Appendix~\ref{app:Cohomology}, these cohomology models can be suitably extended to include a complete set of projectors to produce a stable MBL system for which the entire spectrum of eigenstates take the form of SPT ground-states. The cohomology classification exhausts all known bosonic SPT phases in 2D, but misses a 3D bosonic SPT protected by time-reversal.\cite{AshvinSenthil}  This beyond cohomology state can be characterized by a symmetry respecting surface topological order (STO) that takes the form of a twisted $\Z_2$ gauge theory in which all fractional particles are fermions.  This STO forms the basis of a different set of commuting projector lattice models - Walker-Wang (WW) models\cite{WalkerWang,Burnell2013} - whose uniform ground-state 3D bosonic SPT phases with a given STO. Together, these constructions demonstrate that any bosonic SPT in arbitrary dimension is MB localizable.  

\renewcommand{\arraystretch}{1.2}
\begin{table*}[]
\begin{tabular}{|c|c|c|c|}
\hline 
\bf Type of Phase & \bf Spatial dimension (d) & \bf MBL? & \bf Model type\\
\hline
Chiral Phases & 2D & N & \\ \hline 
\multirow{2}{*}{Boson SPTs} & 1D,2D & Y & cohomology \\
& 3D & Y &cohomology or Walker-Wang \\ \hline
\multirow{2}{*}{Free Fermion SPT} & 1D & Y & \parbox[t]{3in}{\centering noninteracting Kitaev chains} \\
& 2D,3D & N & \\ \hline 
Interacting Fermion SPTs & 2D & Mixed? & ?? \\ \hline 
\end{tabular}
\caption{{\bf MB localizability of various integer topological phases - } (`Type of phase' column) listed according to number of spatial dimensions, $d$, and whether or not the phase is MB localizable (`MBL?' column - `Y/N' indicates Yes/No).  If the phase is MB localizable - an example of the type of commuting projector model is given (`Model type' column).}
\label{tab:Summary}
\end{table*}

\subsection{Free fermion SPTs}
Since noninteracting bosons simply condense to form a gapless superfluid, bosonic SPT phases require strong interactions to obtain an insulating state.  By contrast, a non-interacting fermions can achieve an insulating SPT state simply by filling topological bands.  Local commuting projector Hamiltonians of one-dimensional fermion SPTs are possible purely within such a free fermion description, as famously exemplified by Kitaev's superconducting wire.\cite{Kitaev,MBLQuantumOrder} Generalization of Kitaev's construction may be used to establish the MB localizability of all known one dimensional SPTs with various symmetries.

However, in dimensions larger than one, free fermions filling SPT bands turns out to be a poor starting point for MBL, due to a fundamental obstruction to fully localizing the orbitals of  topological bands. Such obstructions are well known for chiral and time-reversal protected SPT phases, \cite{Thouless1984,Z2WannierObstruction}, and also can be shown to extend (see below) to general fermion SPT bands with discrete unitary symmetry. This obstruction renders topological bands perturbatively unstable to thermalization upon the inclusion of interactions.\cite{RNACP}  

To establish this free fermion obstruction for SPTs with unitary symmetries, consider a finite two- or three-dimensional system of noninteracting fermions with open boundaries described by commuting projector Hamiltonian, (\ref{eq:HCP}), with onsite unitary symmetry group $G$.  Further, take uniform couplings, $\lambda_{\v{r},a} = -\Delta$ for occupied bands ($a=1\dots n_\text{occ}$) and zero for the remaining bands ($a=n_\text{occ}+1\dots N$).  A finite system with open boundaries can be chosen to have highly degenerate set of dispersionless edge states by omitting projectors that intersect the boundary.  We can remove this degeneracy (except perhaps a finite residual degeneracy required by symmetry) by adding noninteracting symmetry preserving edge perturbations of characteristic energy scale $V$. For $V\ll \Delta$, the edge bands reside inside the bulk gap.  By contrast, for a nontrivial fermion SPT, the edge bands of a nontrivial SPT must disperse across the bulk gap and connect to gapped bulk modes in order to inherit the anomalous symmetry properties required for a nontrivial SPT edge state (see Appendix~\ref{app:FreeFermionObstruction} for a formal proof).  Hence, we see that free fermion commuting projector Hamiltonian can only describe trivial (not SPT) phases in dimensions higher than one.  

While these simple arguments rule out the MB localization of higher dimensional \emph{non-interacting} fermion SPTs, they leave open the question of whether \emph{interacting} fermion SPTs can be MB localized, to which we now turn. Adding weak interactions to a non-interacting fermion ``band" SPT is known\cite{RNACP} to further destabilize MB localization and lead to ergodicity and thermalization. Despite this, we will encounter an example in which strong, non-perturbative interaction effects can instead enable MB localization. More generically, we will also give arguments that certain phases cannot be MB localized, even with arbitrary interactions. These results are summarized in Table~\ref{tab:Summary}.

\section{2D Interacting Fermion SPTs}
The key feature that distinguishes fermion systems from purely bosonic ones, is that any local operator contains an even number of fermions - implying that fermion parity is always conserved.  Hence, given a symmetry group $G$, for a fermion system we should really consider a larger ``symmetry" group $G\times \Z_2^F$ (assuming that $\Z_2^F$ is not already contained as a subroup of $G$) - where $\Z_2^F$ is the $\Z_2$-group generated by the fermion parity operator $P_F=(-1)^{n_F}$ and $n_F$ is the fermion number operator. This special property of fermionic systems will turn out to have important implications to their MB localizability. 

In two dimensions, it is then always possible to introduce a $\pi$-flux defect such that fermions obtain a phase of $(-1)$ upon encircling the $\pi$-flux.  In a topological superconductor, this $\pi$-flux is simply a superconducting vortex. For 2D fermion SPTs (f-SPTs), the symmetry properties of the $\pi$-flux provide an important characterization of the phase, and delineate three distinct types of SPT phases:
\begin{enumerate}
\item {\bf Projective f-SPTs - } in which the $\pi$-flux contains symmetry-protected degeneracy associated with fermionic zero-modes on which symmetry acts projectively, and
\item {\bf Fractional Charge f-SPTs -} in which the $\pi$-flux is non-degenerate but carries fractional symmetry charge.
\item {\bf Bosonizable f-SPTs - } in which the $\pi$-flux is invariant under symmetry, 
\end{enumerate}
Starting with a gapless symmetry preserving edge of a bosonizable f-SPT, it is possible to produce an energy gap for all fermion excitations at the edge without breaking symmetry by condensing the $\pi$-flux at the edge (i.e. allow fluxes to freely tunnel in and out of the edge).  This procedure involves strongly fluctuating phase-degrees of freedom - and hence requires interactions. Hence, an interacting bosonizable f-SPTs is topologically equivalent to a purely bosonic SPT accompanied by ``spectator" fermions that form a trivial (Mott) insulator and do not play an essential role in producing the SPT order. Another way to see this is to imagine gauging fermion parity (i.e. coupling fermions to a dynamical $\Z_2$ gauge field). In the gauged theory, the $\pi$-flux is a gapped bulk \emph{quasiparticle}, often called a vison, rather than a confined \emph{defect} (that must be induced by an external field) as in the ungauged theory.  In a gauged bosonizable SPT phase, the vison is bosonic and has trivial symmetry properties - and can be condensed to confine the $\Z_2$ gauge charges (i.e. the fermions) without disrupting the SPT order.  The resulting phase is a purely bosonic SPT phase with no fermion excitations, but with the same SPT properties as the ungauged fermion SPT.

In contrast, fermions play an essential role in establishing the SPT order when the $\pi$-flux has nontrivial symmetry properties.  For projective f-SPTs, the $\pi$-flux has symmetry protected degeneracy associated with bound zero-energy fermion modes.  Consequently - condensing the $\pi$-flux at the edge (or gauging fermion parity and condensing the vison) proliferates these gapless modes and does not produce an energy gap for fermion excitations.  Similarly, in fractional charge SPTs the $\pi$-flux cannot be condensed without breaking symmetry, since the fractional charge of the $\pi$-flux cannot be screened away by the integer charged fermions. 

We will be able to show that all bosonizable f-SPTs are MB localizable with interactions - despite having a free fermion obstruction that prevents their localization in the absence of interactions.  In contrast, projective f-SPT order presents an obstacle for MB localization, which we argue may potentially be fundamental, and present for arbitrary interactions.
At this time, we are not able to make a definitive statement regarding fractional charge f-SPTs. Though exactly solvable models for systems with fractional charge f-SPT ground states have been obtained from the supercohomology construction\cite{Supercohomology}, these models operate on a constrained Hilbert space and it is presently unclear whether this construction yields a complete set of projectors that label all excited states. We leave this as a subject for future work (see also the discussion section below).

\section{Obstacle to MB localizing certain interacting 2D fermion SPTs}
Let us give an example of how a non-chiral topological phase of fermions presents an obstruction to being described by a  Hamiltonian composed of local commuting projectors. We will then extract general lessons for which phases may suffer the same obstruction. 

We consider a simple topological phase of 2D interacting fermions which can be viewed as a condensate of decorated domain walls (as described below) and use this picture to try to construct the commuting projector Hamiltonian. This approach will be shown to fail - which then raises the question of whether this is a fundamental obstruction, or simply the limitation of the approach adopted. We then argue that it is indeed the former, by first showing that a commuting projector Hamiltonian for such a phase necessarily implies that it is a condensate of decorated domain walls, and any such condensate, in the context of the fermion SPT we are considering,  will necessarily produce correlations that decay {\em at least } exponentially. This poses a contradiction since correlation functions of separated local operators must have vanish in the zero correlation length ground states of commuting projector Hamiltonians. 

Consider a 2D topological superconductor with a unitary $\Z_2$ symmetry (in addition to the ever-present fermion parity conservation ``symmetry"), which has a classification labeled by integer topological invariant $\mathcal{N}$. The $\mathcal{N}=1$ SPT phase is obtained by combining opposite chiral superconductors, one a $p_x+ip_y$  and the other a $p_x-ip_y$ superconductors where the $\Z_2$ symmetry acts trivially on the fermions in the $p_x+ip_y$ layers, and gives a phase $(-1)$ to those in the $p_x-ip_y$ layer. The edge consist of a  right and left moving chiral Majorana fermions $\gamma_{s}$ where $s= R,L$ labels right and left movers respectively, described by the continuum Hamiltonian:

\begin{align}
H_\text{edge} = -iv\int dx \gamma_{s}^T\sigma^z_{s,s'}\d_x\gamma_{s'} 
\label{eq:Hedge}
\end{align}
the mass term $H_m= i\int dx \, M\gamma_R\gamma_L$ changes sign under the  $\Z_2$ symmetry and is hence forbidden. As for any $\Z_2$ symmetric state, this $\mathcal{N}=1$ state can be viewed as a string-condensate of $\Z_2$ domain walls in the symmetry-breaking mass, $M$. However, unlike a trivial paramagnet, the domain walls of the $\Z_2$ phase are topologically non-trivial, and can be thought of as being ``decorated" with 1D TSC chains that possess unpaired Majorana end states\cite{Kitaev}. This decoration is revealed by performing a gedanken experiment in which the $\Z_2$ symmetry is broken in two opposite ways in the top and bottom halves of the system. The resulting domain wall terminates at the boundary where it is associated with a change in sign of the mass $M(x)=M_0\cdot{\rm sign}(x)$.  As can be readily verified from Eq.~\ref{eq:Hedge}, such a domain wall carries binds a Majorana zero mode: $\gamma_0 = \int dx \frac{e^{-|x|/\xi}}{\sqrt{2}}\[\gamma_L(x)+\gamma_R(x)\]$, near $x=0$, where $\xi \approx \frac{v}{M_0}$. 

Hence we can view the 2D $\mathcal{N}=1$ TSC state as a fluctuating paramagnetic state of $\Z_2$ ``spins" whose DWs are ``decorated" with 1D Majorana chains. Since the bulk domains have closed boundaries, and the decorating 1D SPT has an energy gap, so too does the bulk of the decorated domain wall (DDW) state. Related DDW constructions have been discussed for bosonic SPT phases\cite{ChenDDW}. Here we attempt an analogous construction for the fermionic 2D TSC, to generate parent Hamiltonians of local commuting projectors. We will find that this fermionic DDW construction suffers a fatal flaw.

\subsection{Explicit DDW construction attempt} 
To implement the DDW construction, we introduce bosonic degrees of freedom that transform as doublets under the $\Z_2$ symmetry.  Specifically, we consider Ising spins $\tau^z_P = \pm 1$ that reside on plaquettes, $P$, of a honeycomb, and complex fermions with annihilation operators $c_{r}$ on the sites $r$ of the honeycomb. The $\Z_2$ subgroup of the symmetry flips the Ising spins but does not effect the fermionic degrees of freedom. 

A protypical $\Z_2$ symmetric state can be expressed as a superposition over all Ising spin-configurations, or equivalently as a equal superposition of domain wall (DW) configurations $\mathcal{C}$ that divide $\tau^z=\pm 1$ domains:
\begin{align}
|\Psi_\text{PM}\> &= \frac{1}{\sqrt{N_\mathcal{C}}} \sum_{\mathcal{C}}|\tau[\mathcal{C}],\psi[\mathcal{C}]\>
\end{align}
where $\{|\psi[\mathcal{C}]\>\}$ are DW-configuration dependent wavefunctions for the fermionic degrees of freedom.  We can produce a nontrivial 2D SPT phase by choosing $|\psi[\mathcal{C}]\>$ to a state in which fermions residing on a $\Z_2$ domain wall form a 1D TSC chain, while fermions residing inside ordered $\Z_2$ domains form a trivial insulating state.

This ``decorated domain wall" (DDW) approach has previously been used to obtain (non-commuting projector) lattice models of bosonic SPTs.\cite{ChenDDW}  Here, we generalize this construction by decorating Ising DWs with 1D fermionic SPTs, and by modifying the terms that proliferate DWs to obtain a commuting projector model.  

The Hamiltonian consists of projectors that decorate DWs of $\tau_P^z$ with Kitaev chains, and transverse field terms that flip $\tau_P^z$ and rearrange the fermions to preserve the DW decorations. The domain wall decoration is accomplished by:
\begin{align} H_\text{decorate} &= \sum_{\mathcal{C}}\Pi_\mathcal{C}H_{F}(\mathcal{C})
\label{eq:HDecorate}
\end{align}
where $\sum_{\mathcal{C}}$ indicates a sum over all $\tau$ configurations $\mathcal{C}$, and $H_F(\mathcal{C})$ takes the form of a zero correlation length Hamiltonian of a Kitaev TSC chain for fermion links residing on the domain walls of $\tau$ and is the Hamiltonian of a trivial product state insulator for sites within ordered domains of $\tau$. To explicitly write the decoration terms, it is convenient to represent the complex fermions by Majorana fields $\chi_{r} = c_{r}+c_{r}^\dagger$ and $\eta_{r}=-i\(c_{r}-c_{r}^\dagger\)$, such that $c_{r}=(\chi_{r}+i\eta_{r})/2$, then: 
\begin{align}
H_F(\mathcal{C}) = -\sum_{r\in \text{DW}}i\eta_{r+d}\chi_r-\sum_{r\notin\text{DW}}i\chi_r\eta_r
\label{eq:HKitaevDecoration}
\end{align} 
Here, one needs to choose an (arbitrary) orientation convention for the DWs, for example the DWs can be oriented clockwise around the $\tau^z=+1$ domains.

%
%

\begin{figure}[t!]
\begin{center}
\includegraphics[width = 1.5in]{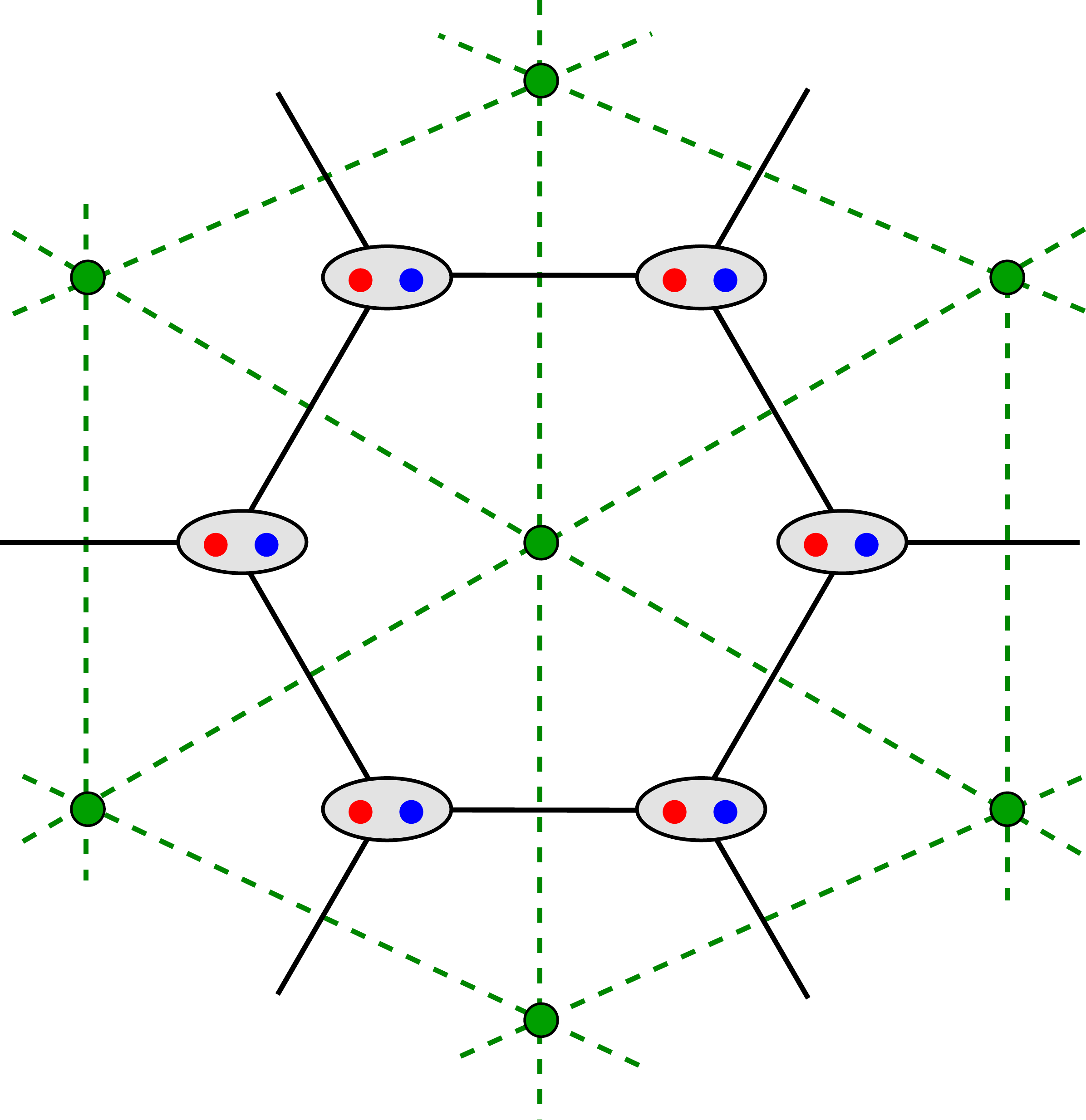}
\vspace{-.2in}
\end{center}
\caption{{\bf Schematic representation of fermionic DDW lattice model - }  The model consists of Ising spins, $\tau$ sit on green sites, which form triangular lattice, and complex fermions, $c_{\v{r}}$ at sites, $\v{r}$ of a honeycomb, which may be decomposed into two real Majorana fermions $\chi$ (red dot) and $\eta$ (blue dot).
}
\label{fig:Lattice}
\end{figure}

\begin{figure}[t]
\begin{center}
(a) \includegraphics[width = .8in]{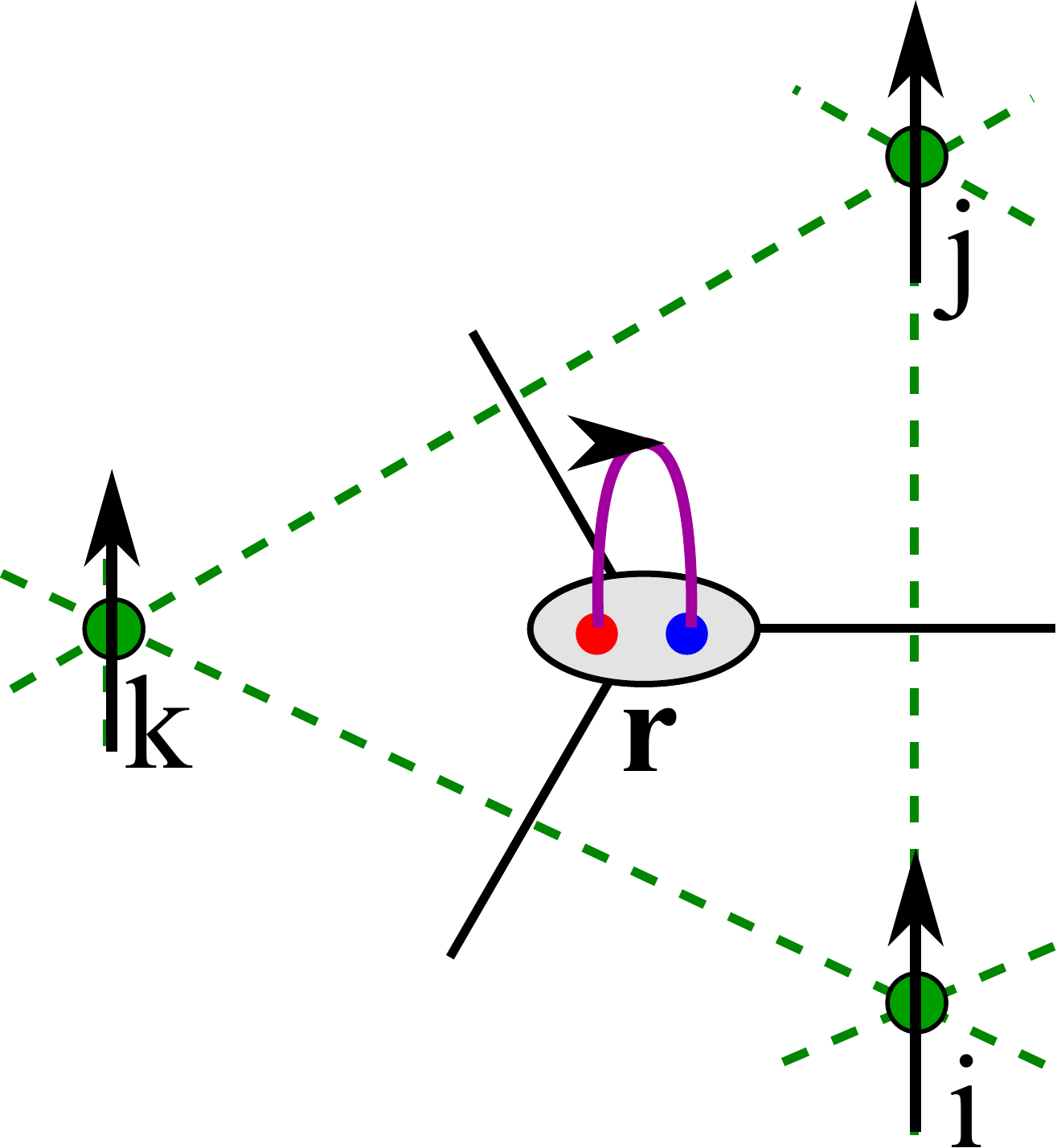}
(b) \includegraphics[width = .9in]{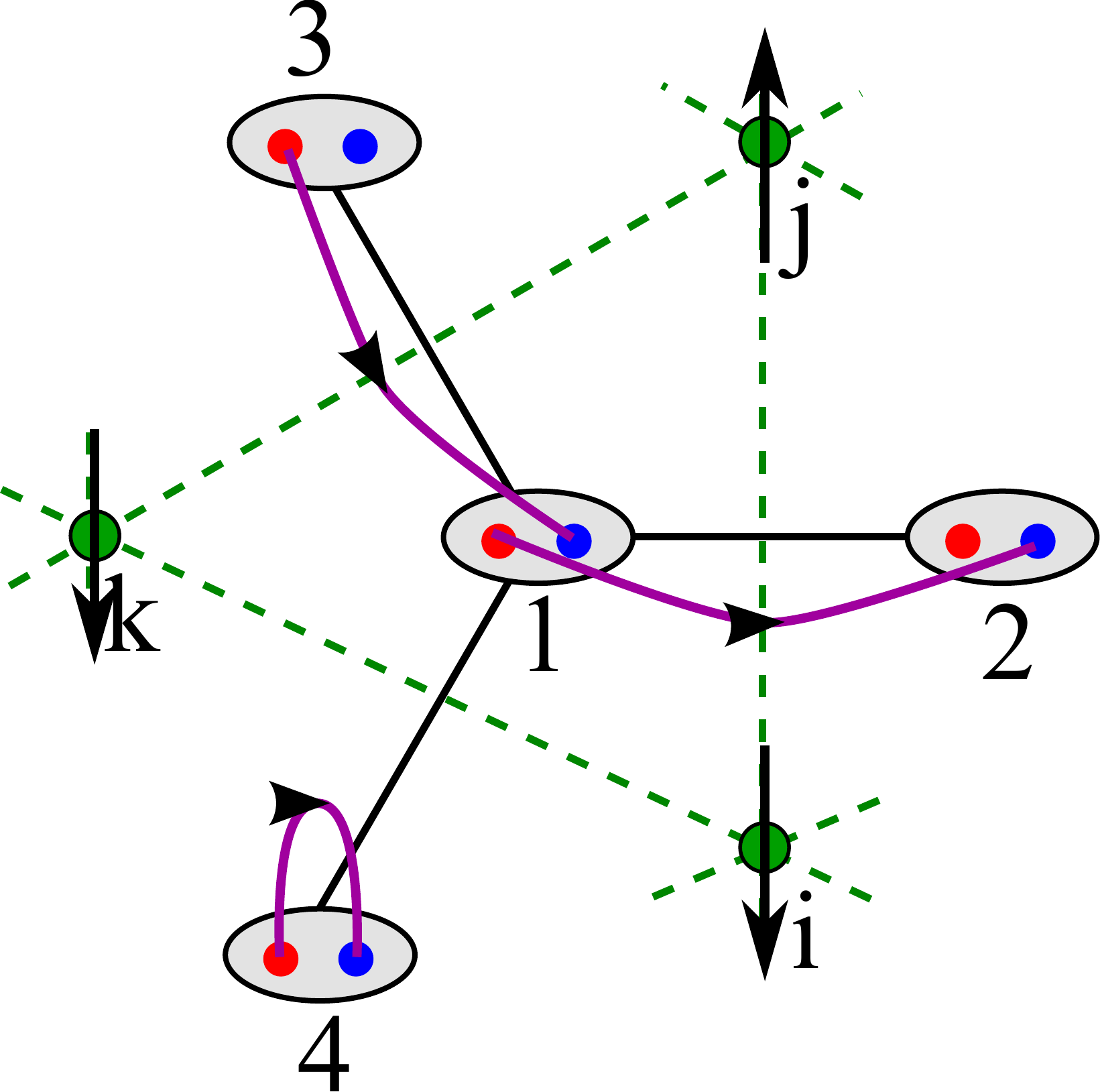}
(c) \includegraphics[width = .9in]{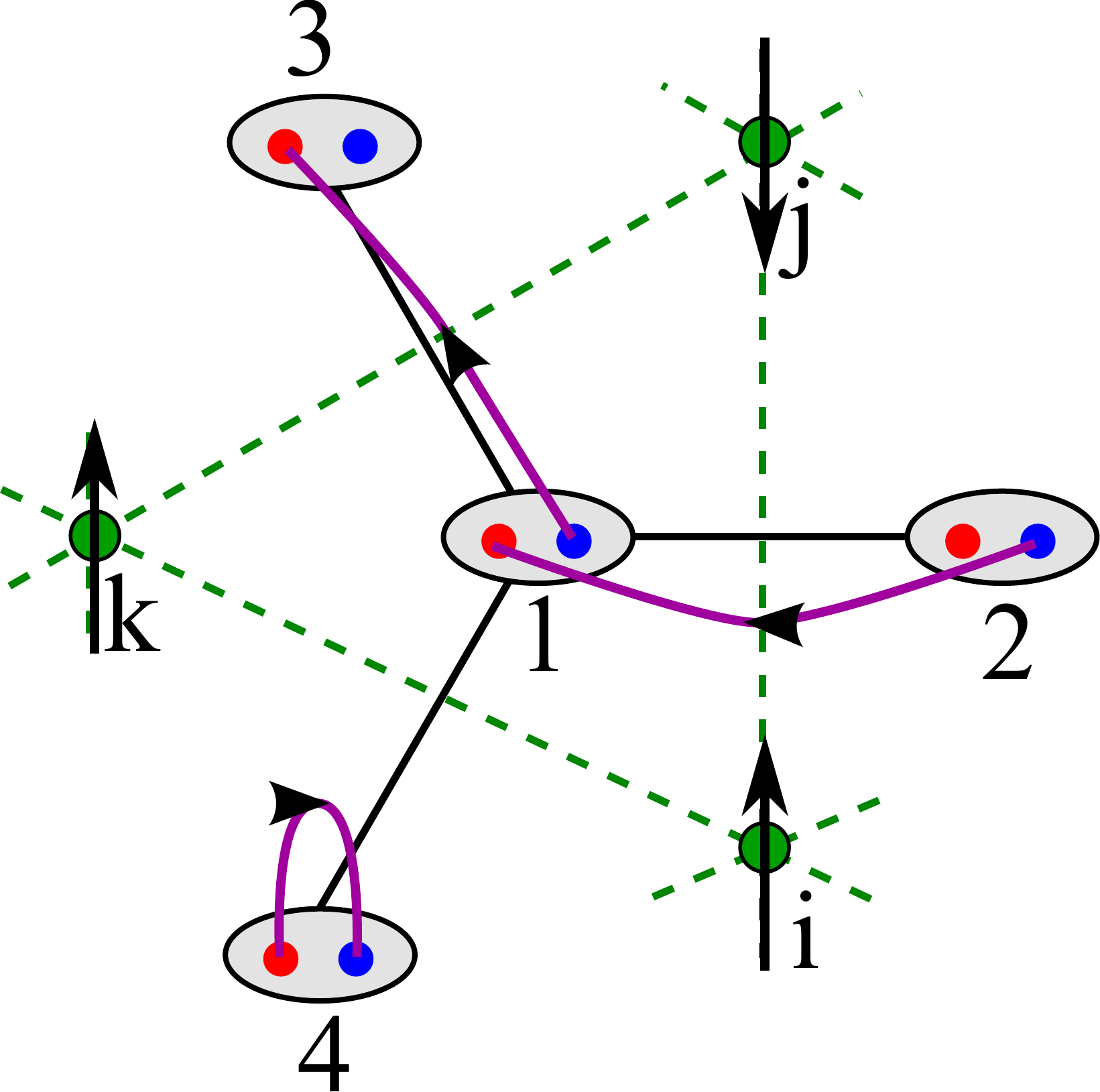}
\end{center}
\caption{{\bf Domain wall condensation (``transverse field") terms - } (a) Projector with no DWs. (b,c) Schematics of selected projectors that attach Kitaev chains to DWs. Purple link with arrow from $\chi_r$ to $\eta_{r'}$ indicates a $i\chi_r\eta_{r'}$ term for fermions.
}
\label{fig:FermionSite}
\end{figure} 

To restore the $\Z_2$ symmetry, one needs to proliferate DWs in the $\tau$ spins, while preserving the DW decoration.  This is accomplished by the generalized ``transverse-field" terms:
\begin{align}
H_\text{TF} &=  \sum_{P}\sum_{\mathcal{C}}\tau_P^x~\Pi_{\mathcal{C}} |F_{\tilde{\mathcal{C}}(P)}\>\<F_{\mathcal{C}}|
\label{eq:HTF}
\end{align}
Here, $\sum_{\mathcal{C}}$ indicates a sum over all $\tau$ configurations $\mathcal{C}$, $|F_{\mathcal{C}}\>$ is the fermionic wave-function corresponding to the ground-state of Eq.~\ref{eq:HDecorate} with $\tau$'s in configuration $\mathcal{C}$, and $\tilde{\mathcal{C}}(P)$ is the spin configuration obtained from $\mathcal{C}$ by applying $\tau_P^x$ to the spin on plaquette $P$.  Importantly, despite superficially appearing to depend on the global spin configuration $\mathcal{C}$, for fixed $P$, the generalized transverse-field terms in Eq.~\ref{eq:HTF} actually involve only a bounded set of operators in the vicinity of $P$.  This is because the ground state of the decoration terms in Eq.~\ref{eq:HDecorate} are determined by purely local rules, and hence $|F_{\tilde{\mathcal{C}}(P)}\>\<F_{\mathcal{C}(P)}|$ involves only those fermionic degrees on the sites bordering plaquette $P$.

\subsubsection{Fermion Parity Obstruction}
While the above construction appears to yield an exactly solvable lattice model of a 2D interacting TSC, it actually contains a fatal flaw. Specifically, a complication arises due to the fact that the ground-state of a single Kitaev superconducting chain with periodic boundary conditions has odd fermion parity compared to the trivial ground state. To see this, consider the fermion parity operator along a closed loop of sites $\Gamma$: $P_F = \prod_{r\in \Gamma} i\chi_r\eta_r = -\prod_{r\in \Gamma}i\eta_r\chi_{r+1}$. In the trivial ground state, $i\chi_r\eta_r=1$ for each site, and in the 1D TSC ground state with periodic boundary conditions along $\Gamma$, $i\eta_r\chi_{r+1}=1$, demonstrating that $P_F$ is opposite for the trivial and topological loops. This seemingly innocuous observation will turn out to present a serious obstacle to MB localizing the $\mathcal{N}=1$ phase.

Clearly, the parity switching feature presents a problem for the DDW construction with $\mathcal{N}=1$ since, here, the desired TF operators $|F_{\tilde{\mathcal{C}}(P)}\>\<F_{\mathcal{C}}|$ actually change fermion parity whenever ${\tilde{\mathcal{C}}(P)}$ and $\mathcal{C}$ differ by an odd number of domain walls.  Such fermionic operators are highly nonlocal (they anticommute with other fermion operators even arbitrarily far from plaquette $P$) and hence are not permitted in a local Hamiltonian.  At this point we may conclude that perhaps the DDW approach is not inadequate in this particular situation or may need to be modified e.g. by considering condensates of domain walls that enclose $\pi$ flux, and hence even fermion parity. We show below that neither of these is a viable option. First we argue that a condensate of $\pi$ flux domain walls has correlations that decay at least exponentially with distance. Such correlations should be absent in ground states of local projectors. On the other hand, we argue that such a condensate is mandated by the assumption of a parent Hamiltonian of local commuting projectors, which forces the desired contradiction.

\subsubsection{Finite Correlations in the Decorated Domain Wall State}
\begin{figure}[tb]
\begin{center}
\includegraphics[width = 3in]{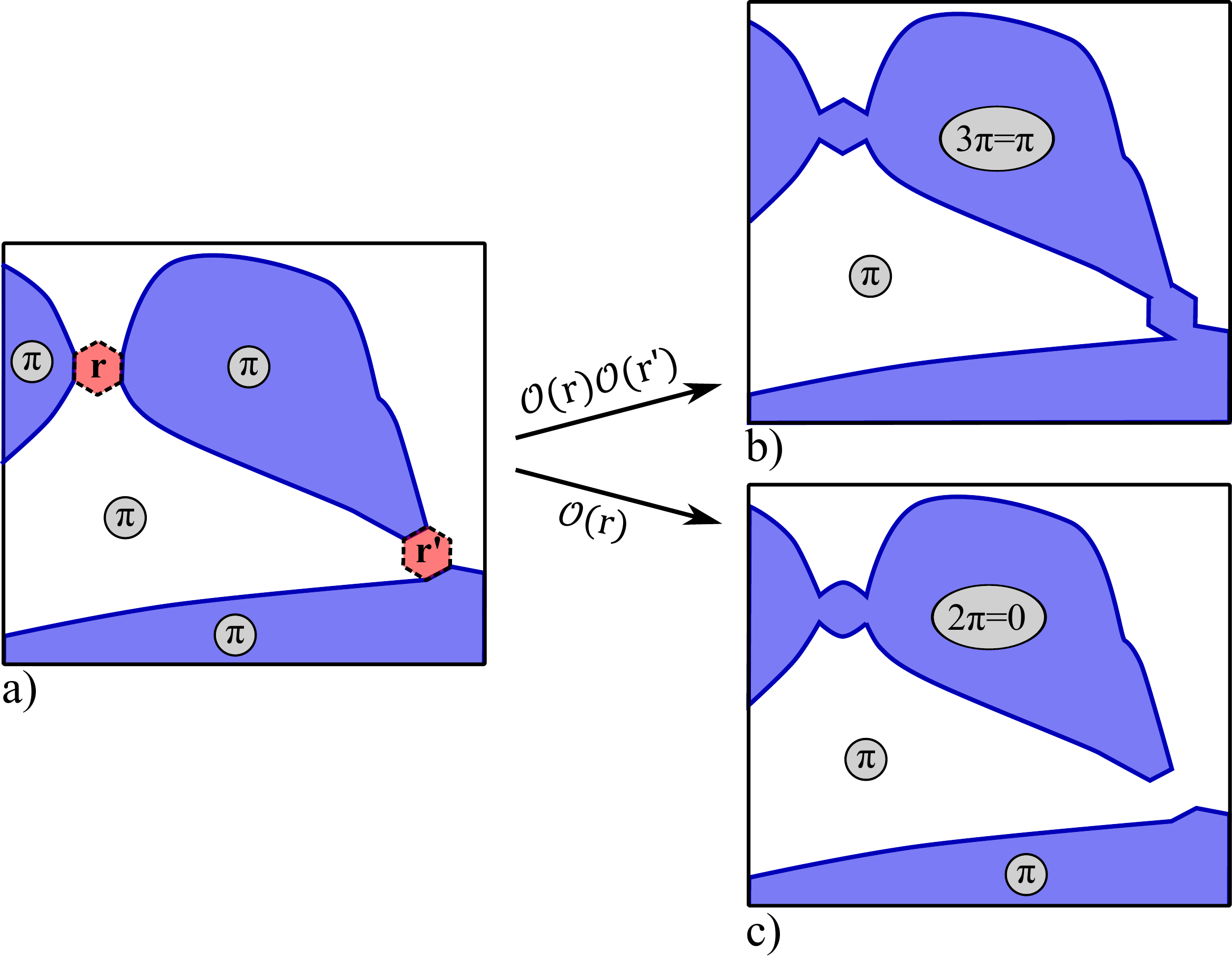}
\end{center}
\caption{{\bf Non-zero correlation length - } A schematic illustration of a DW configuration, $\mathcal{C}$, for which the correlation function of the flux-less loop insertion operator $\mathcal{O}$ fail to factorize - implying non-zero correlation length and an obstacle to constructing a strictly local projector model.
}
\label{fig:CorrelationLength}
\end{figure}

The difficulty of  nontrivial fermion parity of the Majorana chains decorating the domain walls can be circumvented by twisting the boundary conditions to add a $\pi$-flux Eq.~\ref{eq:HTF}, but we will see that restoring locality to the DDW Hamiltonian in this way comes at the expense of spoiling the exact solvability of the model. To see this, consider the operator $\mathcal{O}(\v{r})$ acting on a minimal plaquette at position $\v{r}$ that adds a flux-free domain with appropriately decorated domain wall.  Specifically, acting on $|\psi[\mathcal{C}]\>$, $\mathcal{O}(\v{r})$ either creates a fluxless loop if the plaquette containing $\v{r}$ resides inside a connected domain of the configuration $\mathcal{C}$, or grows an existing domain (without adding additional flux), if the plaquette borders a domain of $\mathcal{C}$ (see Fig.~\ref{fig:CorrelationLength}).

The expectation value of two such operators $\<\mathcal{O}(\v{r})\mathcal{O}(\v{r}')\>$ does not factorize into $\<\mathcal{O}(\v{r})\>\<\mathcal{O}(\v{r}')\>$ even for widely separated $\v{r},\v{r}'$. To see this, consider the loop configuration, $\mathcal{C}$, shown in Fig.~\ref{fig:CorrelationLength}. Acting on this configuration $\mathcal{O}(\v{r})$ merges two domains (each with $\pi$-flux) without adding any flux - producing a domain with $2\pi\simeq 0$ flux. Such fluxless domains are not present in the ground-state $|\Psi_\text{2DSPT}\>$, implying $\<\Psi_\text{2DSPT}|\mathcal{O}(\v{r})|\psi[\mathcal{C}]\>=0$.  On the other hand, acting on $|\psi[\mathcal{C}]\>$ with both $\mathcal{O}(\v{r})$ and $\mathcal{O}(\v{r}')$ merges three domains together - and hence preserves the odd flux through the domains, implying: $\<\Psi_\text{2DSPT}|\mathcal{O}(\v{r})\mathcal{O}(\v{r'})|\psi[\mathcal{C}]\>\neq 0$.

In general, 
\begin{align}
C(r,r')&=\<\mathcal{O}(\v{r})\mathcal{O}(\v{r}')\>-\<\mathcal{O}(\v{r})\>\<\mathcal{O}(\v{r}')\>\approx 
\frac{N_{\mathcal{C}}'}{N_\mathcal{C}}\approx e^{-|\v{r}-\v{r}'|/\xi}
\end{align}
where $N_\mathcal{C}'$ is the number of configurations of the sort shown in Fig.~\ref{fig:CorrelationLength}.  These special configurations require $\v{r}$ and $\v{r'}$ to be connected by a single symmetry domain. Since the SPT ground-state (by definition) must preserve symmetry. Then, the ratio $\frac{N_{\mathcal{C}}'}{N_\mathcal{C}}$ is proportional to the probability that a single domain percolates from $\v{r}$ to $\v{r'}$. In any gapped symmetry preserving phase, this probabiltiy decays exponentially with separation $|\v{r}-\v{r}'|$ with a non-zero characteristic lengthscale $\xi$. This non-zero correlation length, $\xi$, contradicts the assumption that the phase can be described by a strictly local commuting projector Hamiltonian whose correlation functions factorize exactly beyond some finite lengthscale. We note that alternative methods for circumventing this fermion parity obstruction, e.g. by adding extra fermionic degrees of freedom to the DW loops, are topologically equivalent to condensing $\pi$-flux loops and should suffer from the same non-locality.

These considerations strongly suggest that the requirement that each symmetry domain contain a $\pi$-flux is incompatible with the commuting projector form of the Hamiltonian, and that, consequently, there is a fundamental obstruction to deforming this $\mathcal{N}=1$ TSC to the strictly localized limit. This obstruction is present for any projective f-SPT where the 1D SPT decorating the symmetry domain wall is topologically equivalent to a collection of Kitaev TSC chains. 

A potential shortcoming of the above argument is that we assume that it is possible to construct a flux-less insertion operator, $\mathcal{O}$, that has non-vanishing two-point functions in some eigenstate. While there is no symmetry principle requiring $\<\mathcal{O}(\v{r})\mathcal{O}(\v{r}')\>$ to vanish in all eigenstates, it is conceivable that one may be able to construct a highly fine-tuned model for which this two-point function vanishes in every eigenstate, possibly enabling a loop hole in our argument.

\subsubsection{Necessity of the DDW perspective for $\mathcal{N}=1$}
At this point, one might suspect that the above problems merely demonstrate a shortcoming of the DDW approach for building solvable lattice models of fermionic SPT phases. However, we will argue that the DDW description and its associated problems for creating a commuting projector Hamiltonian are fundamental to the $\mathcal{N}=1$ state, and therefore unavoidable. 

We first show that, given a putative local commuting projector Hamiltonian for the $\mathcal{N}=1$ TSC with $\Z_2$ symmetry, we can formally construct a family of DDW insertion operators that commute with the Hamiltonian. For the reasons outlined above, this then suggests a non-vanishing correlation length in contradiction with the assumption of a local commuting projector Hamiltonian.

A DDW insertion operator acts with the $\Z_2$ symmetry generator inside a closed connected region, $A$, creating an extra $\Z_2$ domain, and also inserts a Kitaev chain along the boundary. To define a local action of symmetry consider acting with the generator of the $\Z_2$ symmetry, $g$, in a bounded region, $A$, implemented by the operator: $U_A(g) = \prod_{i\in A} g_i$, where $g_i$ is the generator of group element $g$ acting on physical site $i$.  Under the action of $U_A(g)$ the original projectors $\{\Pi\}$ map to a new set of commuting projectors: $\tilde{\Pi}_{\v{r},a} =  U_A(g)\Pi_{\v{r},a} U_A(g)^\dagger$ (note that $\{\tilde{\Pi}\}$ are also a commuting set of projectors). Projectors residing in the interior of $A$, and in its complement $A^c$ are unchanged, but those whose support intersects the boundary of $A$, $\d A$ are generically altered.  However, we may repair projectors overlapping $\d A$ by acting a string operator, $W_{\d A}(g)$, with bounded support in the vicinity of $\d A$, that maps the projectors transformed by $U_A(g)$ back to the original projectors. Thus, for any local commuting projector Hamiltonian, we may ``pull back" the local action of symmetry on a 2D domain, into a 1D string operator on its boundary.

\begin{figure}[tb]
\begin{center}
\includegraphics[width = 3in]{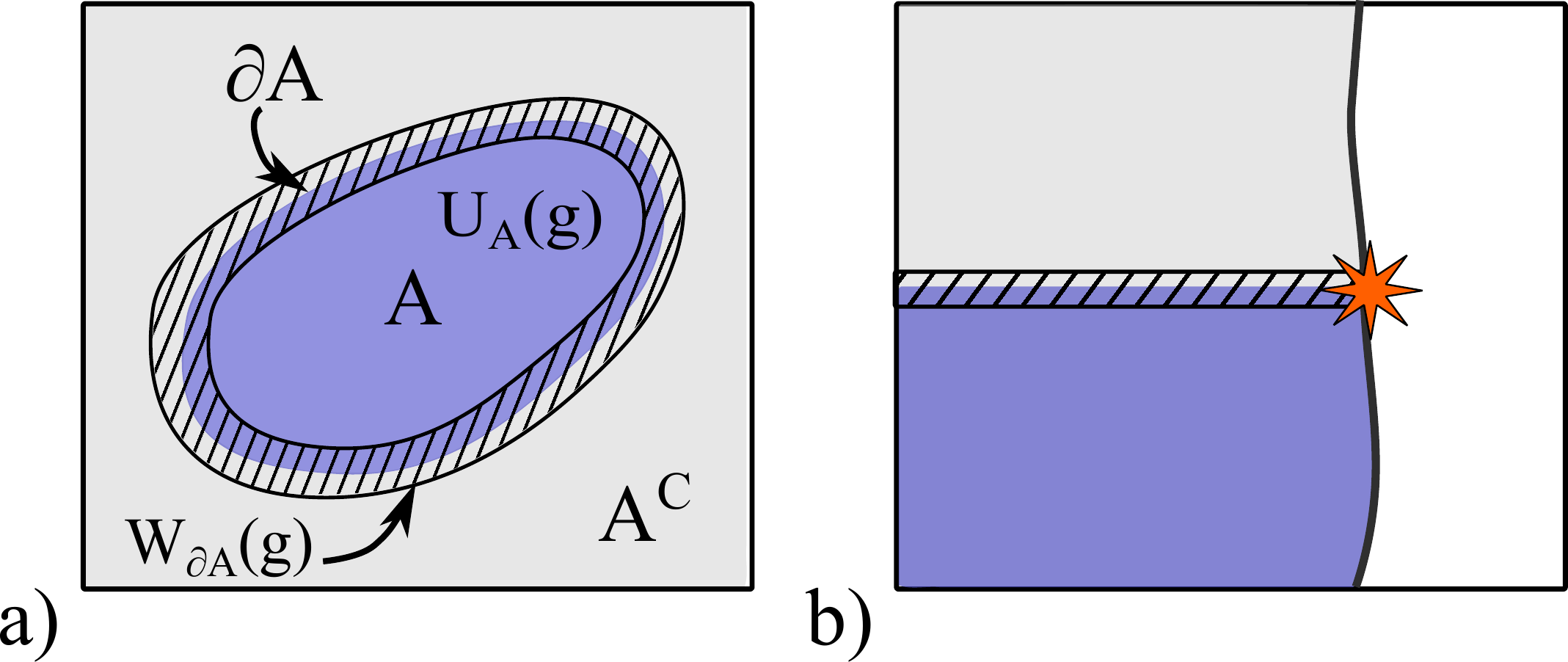}\\
c) \includegraphics[width = 1.3in]{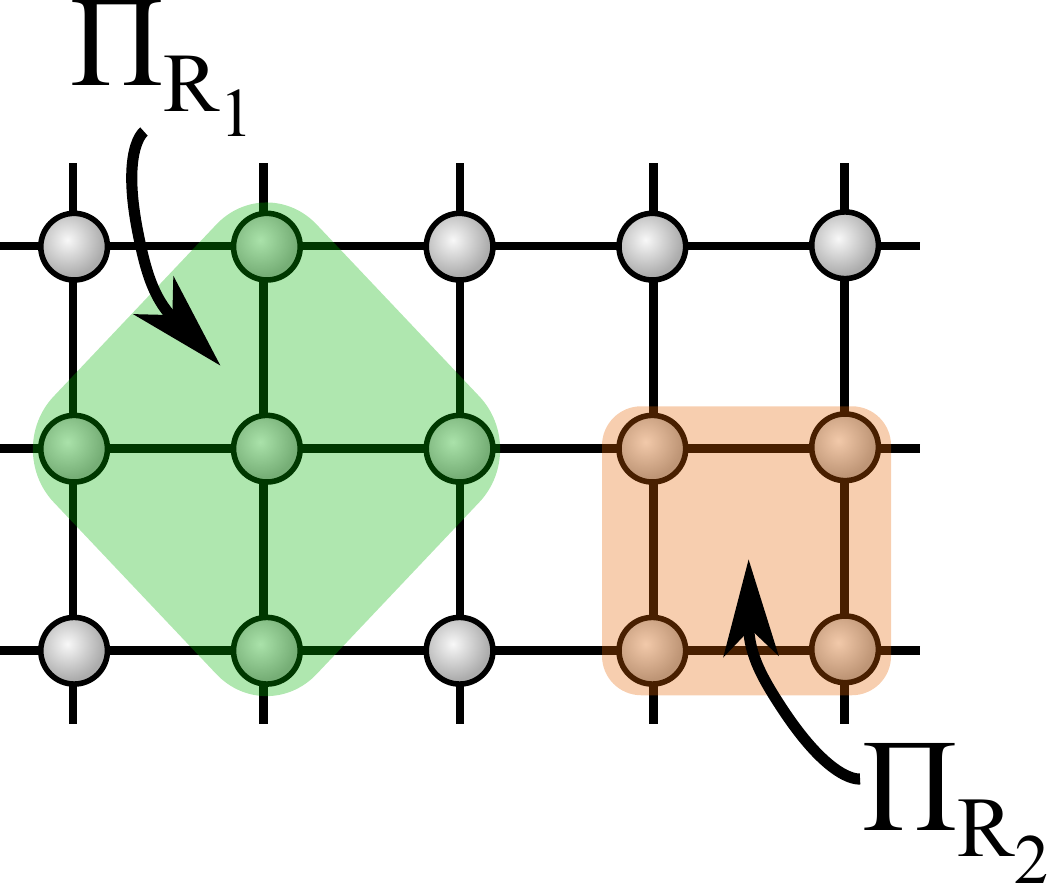}
\end{center}
\caption{{\bf Formal construction of DDW insertion operator - } (a) Given a commuting projector Hamiltonian, one can construct an operator that commutes with the Hamiltonian and inserts a symmetry domain - i.e. which acts with a symmetry group generator, $g$, inside a region $A$ (blue globular region), and a string operator, $W_{\d A}$, on its boundary, $\d A$ (hatched strip). (b) By examining the end of the operator $W_{\d A}$ where it intersects a spatial boundary (orange star), and using the bulk-boundary correspondence, one can show that for projective fermionic SPTs, $W_{\d A}$ inserts a 1D SPT. (c) Though independent, projector terms in the Hamiltonian (schematically indicated by green and orange squares) generically overlap multiple physical sites (gray circles).
}
\label{fig:FermionSite}
\end{figure}

This formal construction yeilds an infinite family of DW insertion operators, $\{D_A(g) = W_{\d A}(g)U_A(g)\}$, which commute with the Hamiltonian.  Hence, the ground-state (indeed all eigenstates) are eigenstates of $\{D_A(g)\}$, and must be regarded as ``condensates" of these DW insertion operators.

To complete the DDW interpretation, let us consider the end of string operator $W_{\d A}(g)$, formally obtained by acting with $D_A(g)$ on a semi-infinite system in the left half plane ($\{x<0,y\in\mathbb{R}\}$) with open boundary along $x=0$, and open boundaries, and take $A$ to be the lower half-plane ($y<0$). By construction, the projectors that do not overlap the point $(x=0,y=0)$, are left unchanged, the operator $D_A(g)$ only effects degrees of freedom near $x=0,y=0$ and thus creates a low-energy excitation on the boundary, i.e. $D_A(g)$ is the boundary operator that inserts a $\Z_2$ symmetry twist (possibly dressed by some local boundary operators at the origin).

Acting on a vortex ($\pi$-flux) in the $\mathcal{N}=1$ 2D TSC, the fermion parity operator anticommutes with the $\Z_2$ symmetry generator, i.e. $P_F D_A(g) P_F = -D_A(g)$.  Since $P_F$ commutes with the local action of $\Z_2$, this implies that $P_F W_A(g) P_F = -W_A(g)$ - i.e. that the 1D SPT decorating the ``domain-wall" of the $\Z_2$ symmetry develops odd fermion parity when encircling a $\pi$-flux.  We have already seen that a loop of 1D TSC, has precisely this property, in fact, this property is sufficient to define the 1D TSC phase - in the sense that any 1D state with this property is topologically equivalent to that described by (\ref{eq:HKitaevDecoration}). This shows that in any putative commuting projector Hamiltonian, one can formally construct DDW insertion operators that commute with the Hamiltonian, i.e. that the DDW interpretation of the $\mathcal{N}=1$ state is indispensable, and that the concomitant barriers to MB localization encountered in the previous section are potentially fundamental. Moreover, these arguments readily extend to other projective f-SPTs, which have similar DDW interpretations and face a common barrier to MB localization.

\subsection{DDW Construction for $\mathcal{N}>1$}
While the DDW viewpoint is not strictly necessary for $\mathcal{N}>1$, we may still attempt to construct solvable lattice models for these phases using the DDW method by extending the above model to include $\mathcal{N}>1$ flavors of fermions, $c_{\v{r},a}$, labeled by flavor index $a=1\dots \mathcal{N}$. However, as there is only one kind of non-trivial 1D TSC wire, to implement the DDW construction for $\mathcal{N}>1$, we must artificially extend the the symmetry group $\Z_2\rightarrow \Z_2\times G$, where $G$ is a (to be specified) internal, unitary symmetry. We can then decorate the $\Z_2$ DWs by 1D SPTs protected by symmetry group $G$.

At first glance, the odd fermion parity problem for $\mathcal{N}=1$ does not affect the DDW models with even number of fermion flavors $\mathcal{N}$; however, a related problem emerges for $\mathcal{N}=2$.  To see this note that, for $\mathcal{N}=2$ terms such as $c_{r,1}c_{r,2}$ must be symmetry forbidden, otherwise it would be possible to deform the DW decoration into a trivial insulator without breaking symmetries. Equivalently, the two different species of fermions, $a=1,2$ must have distinct charges under the residual symmetry group, $G$, of the DW (recall that the DW configuration breaks the $\Z_2$ subgroup of the total symmetry group $\Z_2\times G$).  Suppose we decorate the Ising DWs with 1D fermion SPTs having periodic boundary conditions, as required to obtain a zero-correlation length model.  Then the requirement that $c_{r,a}$ transform differently under $G$ for $a=1,2$, implies that $|F_{\tilde{\mathcal{C}}(P)}\>\<F_{\mathcal{C}}|$ have nontrivial $G$-charge whenever ${\tilde{\mathcal{C}}(P)}$ and $\mathcal{C}$ differ by an odd number of domain walls.

%
%

For $\mathcal{N}=4$, however, such obstructions need not occur. One suitable option is to take $G=\Z_2\times\Z_2$, and give each of the four flavors a different pair of charges from the set from the set $\{(+,+),(+,-),(-,+),(-,-)\}$ under the two $\Z_2$ subgroups of $G$.  For this choice, each of the four copies of Kitaev chains decorating the Ising domain walls has distinct transformation properties, ruling out the bilinear mass terms $c_{r,a}c_{r,b}$.  However, the TF operators $|F_{\tilde{\mathcal{C}}(P)}\>\<F_{\mathcal{C}}|$ all have trivial symmetry quantum numbers as the product of charges of all four fermion species is trivial.  Indeed, for this specific choice, one can readily verify that the DDW Hamiltonian, (\ref{eq:HDecorate},\ref{eq:HTF}), produce the desired symmetry-preserving commuting projector form (see Appendix~\ref{app:N4Model} for details).

Interestingly, the obstruction to localization is lifted precisely because the fermion phase becomes bosonizable at $\mathcal{N}=4$. The underlying bosonic character of the phase obtained from the DDW construction with $\mathcal{N}=4$ can be seen by adding interaction terms $\prod_{a=1}^4\chi_{r,a}$ and $\prod_{a=1}^4\eta_{r,a}$ to the edge of the system.  These gap out all fermionic excitations at the edge, but leave symmetry protected gapless bosonic modes (see Appendix~\ref{app:InteractingEndStates}).  Alternatively, one can examine a $\pi$ flux for the fermions and verify that it has trivial symmetry properties (see Appendix~\ref{app:InteractingEndStates}).  Hence, we could remove the fermions entirely from the problem without effecting the SPT order by coupling the fermions to a $\Z_2$ gauge field and condensing the gauge-flux to confine the fermionic excitations. In this construction, the choice of symmetry implementation that enables the DDW construction to work, also  allows the interaction terms $\prod_{a=1}^4\chi_{r,a}$ and $\prod_{a=1}^4\eta_{r,a}$ that can gap out the fermion degrees of freedom at the edge - producing an essentially bosonic SPT.  This construction then suggests that whether a fermion SPT is bosonizable and whether it is MB localizable are related issues.

\section{Generalizations and discussion}
The above-described obstruction appears to rule out MB localization of 2D projective f-SPTs (DDW phases), and highlights a sharp distinction between bosonic SPT phases which can all be MB localized and permit exact TPS descriptions, and fermion SPTs for which can face obstructions to MB localization unless they are bosonizable. 

At this time, we are unable to decisively state whether the fractional charge f-SPT phases also resist MB localization. The simplest scenario is that all intrinsically fermionic (i.e. not bosonizable) SPT phases share a common obstacle to MB localization, i.e. that the special properties of fermion parity ``symmetry" make it impossible to endow $\pi$-fluxes (or flux lines in 3D) with anomalous symmetry properties. Further suggestive evidence for this scenario comes from considering Walker-Wang type loop models for 3D fermion SPTs. Unlike their bosonic counterparts, which give an MB localizable family of projector Hamiltonians, fermion Walker-Wang constructions actually produce a gauged version of the fermionic SPT order in which the fermions are emergent anyonic particles coupled to a dynamical $\Z_2$ gauge field. These loop models are only symmetry preserving in the zero-gauge-flux sector - since for a fermionic SPT the $\Z_2$ gauge flux lines ($\pi$-flux lines for the fermions) are necessarily gapless or break symmetry. Moreover, while it is possible ``ungauge" these models by coupling them to local fermion degrees of freedom\cite{FermionWW}, this ungauging action spoils the locality of the terms in the Walker-Wang model and does not produce a local commuting projector Hamiltonian. The failure of these Walker-Wang constructions due to the special properties of $\pi$-flux lines in fermion SPTs is consistent with a scenario in which fermion SPTs are intrinsically not MB localizable.

However, the supercohomology results of Ref.~\onlinecite{Supercohomology} suggests that the situation may be more complicated than a simple dichotomy between fermionic and bosonic phases, i.e. that only certain types of fermionic phases (e.g. projective f-SPTs) face an obstruction to MB localization. In particular, exactly solvable models with a fermionic SPT \emph{ground states} (specifically a fractional symmetry charge 2D f-SPTs) was constructed by supercohomology methods in Ref.~\onlinecite{Supercohomology}. While it is conceivable that this construction may be extended (e.g. along the lines of Appendix.~\ref{app:Cohomology}) to obtain an MB localizable Hamiltonian for all \emph{excited} eigenstates, an explicit demonstration is absent. An added complication compared to the bosonic case is that the fermionic models operate in a restricted Hilbert space, which may present complications for constructing a generically MB localizable model. These supercohomology results suggest further subcategorization within fermion SPT phases -- in which only projective f-SPTs suffer a barrier to MB localization. 

A third possible scenario is that only phases like the $\mathcal{N}=1$ 2D topological superconductor phase, in which symmetry defects carry unpaired Majorana zero-modes, suffer an obstacle to localization. Distinguishing among these scenarios will be an important subject for future work.

A second limitation of our current work is that we have so far considered only SPT phases protected by discrete, unitary symmetry groups. Our arguments against MB localization in projective f-SPTs can likely also be extended to antiunitary time-reversal symmetry, since, for local commuting projector Hamiltonians it is possible to go to a tensor-product state description where one can define the local action of an antiunitary symmetry.\cite{GaugingTR} Extending these ideas to SPT phases protected by continuous symmetry group, however, appears problematic. For example, in 2D there is an integer quantum Hall phases of bosons protected by a $U(1)$ charge conservation symmetry.\cite{BosonIQH,CSClassification} The group cohomology approach also produces exactly solvable lattice Hamiltonians for these phases. However, these models utilize $U(1)$ rotor degrees of freedom with an unbounded Hilbert space on each site. The unbounded character of the on-site Hilbert space, inevitably allows for resonant transitions for any given pair of sites somewhere in the spectrum. A related conjecture, that continuous symmetry SPT phases resist localization, was posed for one-dimensional spin chains with continuous rotation symmetry \cite{AnushyaMBLSPT}. Further support for this hypothesis for the case of continuous non-Abelian symmetries was obtained by more systematic analysis\cite{HeisenbergRSRG}. Whether the infinite on-site Hilbert space can be truncated without spoiling the exact solvability of the model, or alternatively whether the accidental resonances in the rotor models are sufficiently prevalent to completely spoil localization are subtle questions that we leave for future work. 

We remark, however, that, if possible, MB localization of these bosonic IQH states would have rather striking consequences: since the $U(1)$ degree of freedom cannot be spontaneously broken at the 1D boundary of a 2D system, due to Mermin-Wagner arguments, bulk localization of the bosonic integer quantum Hall phases could topologically guarantee the presence of gapless, ergodic boundary modes. This situation would be unlike that for discrete-symmetry SPT phases, whose symmetry can always be spontaneously broken at the edge, allowing a localized non-ergodic edge termination. These considerations hint at interesting connections between MB localizability and Mermin-Wagner type arguments that deserve further exploration.\\


\noindent{\textit{Acknowledgements - }} We thank Y.-M. Lu, M. Zaletel, X. Chen, E. Altman, Z.-C. Gu, and M. Cheng for helpful conversations.  ACP was supported by the Gordon and Betty Moore Foundation's EPiQS Initiative through Grant GBMF4307. AV was supported by NSFDMR 120678.\\

\noindent \textit{Related Work - } During the course of completing this manuscript, a related work\cite{Slagle}, appeared, in which formulates an alternate criterion for full MBL protection of SPT order.


\vspace{.25in}
\appendix 
\section{MB localizable models for bosonic SPTs}
\subsection{Completing the cohomology models\label{app:Cohomology}}
Exactly solvable lattice models with SPT \emph{ground-states}, were constructed in Refs.~\onlinecite{CohomologyPRB,CohomologyScience}  for all bosonic SPT phases classified by group cohomology. - i.e. those with discrete unitary on-site symmetry group $G$.  The models involve an appropriate $d$-dimensional lattice (e.g. triangular for $d=2$), with bosonic site-degrees of freedom $\{|g\>_i\}$, on which $G$ acts as such that $g_1|g_2\> = |g_1g_2\>$.  The construction starts from a trivial Hamiltonian: $H_\text{triv} = -\sum_i |0_i\>\<0_i|$ where $|0_i\> = \frac{1}{\sqrt{|G|}}\sum_{g\in G}|g_i\>$.  The SPT Hamiltonian is then obtained as:
\begin{align}
H_0=-\sum_i \Pi_i = -\sum_i U_i|0_i\>\<0_i|U_i^\dagger
\end{align} 
where $U_i$ are unitary matrices diagonal in the $|\dots g_ig_j\dots\>$ basis, with elements consisting of phases chosen from the group cohomology:
\begin{align}
U_i = \sum_{\{g\}}\frac{\prod_{ijk\in\triangle}\nu_{d+1}(1,g_i,g_j,g_k)} {\prod_{ijk\in\triangledown}\nu_{d+1}(1,g_i,g_j,g_k)} |\{g\}\>\<\{g\}|
\end{align}
where $\nu_{d+1}$ are phases obtained from a particular element (SPT phase) of the cohomology group $\mathcal{H}^{d+1}(G,U(1))$, which involve the $g$-values of nearest neighboring sites of $i$.

While $H_0$ has a unique non-trivial SPT ground-state, the commuting projectors $\Pi_{i,0}$ are not complete for $|G|>2$ (where $|G|$ is the number of group elements in $G$).  This is because there are $|G|-1$ states per site that are orthogonal to $U_i|0_i\>$.  For $|G|>2$, these $|G|-1$ orthogonal states are all annihilated by $H$, implying that the excited states are highly degenerate.  Hence, as written the model is not suitable for MBL.  

Fortunately, this issue can be remedied by extending $\Pi_{i,0}$ to include complete set of projectors such that every state in the spectrum behaves as an SPT ground-state in the same phase.  Let us specialize to a finite Abelian group $G$, where we may conveniently work in the Fourier-transformed basis: $|k_i\> = \frac{1}{\sqrt{|G|}}\sum_{g\in G} e^{ik(g)}|g\>$.  For example $G=\Z_N$ there is a single generator $g$, and all group elements can be written as $g^n$ with $n\in 0\dots N-1$, and $|k_i\> = \frac{1}{\sqrt{|G|}}\sum_{g\in G} e^{2\pi ikn}|g^n\>$, and $k \in \{\frac{j}{N}\}_{j=0,\dots N-1}$..  More generally, any finite Abelian group can be decomposed as a product of a finite number of $\Z_N$'s: $G = \Z_{N_1}\times\dots \Z_{N_m}$.  In this case $k$ will be an $m$-component vector $k\in\{\(\frac{j_1}{N_1},\dots \frac{j_m}{N_m}\)\}_{j_i=1\dots N_i-1}$

The Hamiltonians $H_k = -\sum_i U_i|k_i\>\<k_i|U_i^\dagger$ are each related to $H_0$ by the unit-depth local unitary transformation: $H_k = M(k)H_0M(k)^\dagger$, with $M(k) = \prod_i M_i(k)$, and $M_i(k)=\sum_{g_i}e^{ik(g_i)}|g_i\>\<g_i|$, implying that $H_k$'s ground-state is the same SPT phase as $H_0$'s for each $k$. To see this, note that $M_i(k)$ are diagonal in the $|g_i\>$ basis and hence commute with $U_i$ implying: $H_k = -\sum_i U_iM_i(k)|0_i\>\<0_i|M_i^\dagger(k)U_i^\dagger = M(k)H_0M(k)^\dagger$. 
Moreover, it is easy to check that this unitary transformation preserves the commuting projector structure.  Finally, the projectors are manifestly invariant under the symmetry $G$, since  $|k_i\>$ are eigenstates of the symmetry group generators whse eigenvalues have unit amplitude, and since the $U_i$ are constructed from the cohomology to be $G$-invariant.

Then, the commuting projectors $\Pi_{i,k}$ form a complete set that span all excited states of the system, enabling one to construct MB localized family of fixed point Hamiltonians, Eq.~\ref{eq:HCP}, for each SPT phase in the cohomology classification.

%

\section{Free Fermion Obstruction \label{app:FreeFermionObstruction}}
In the main text, we argued that the symmetry respecting edge of an SPT must exhibit bands that disperse across the bulk gap, and that, since commuting projector Hamiltonians produce flat edge bands and since symmetry cannot be broken in a non-interacting fermion system it should be impossible to have a MB localizable SPT of noninteracting fermions. To formalize this intuition let us consider first the case of a 2D SPT with an on-site, unitary symmetry group $G$. This argument readily extends to 3D, by considering one dimensional cuts of the 2D boundary Brillouin zone.  The most general non-interacting SPT edge contains equal number $N$, of left and right moving Majorana fermion modes $\gamma_{R/L,a}$ with $a\in\{1\dots N\}$ (complex fermion zero modes are special case of $N$ even).  The left and right movers must cross at some point within the Brillouin zone, and without loss of generality we may assume that the crossings occur momentum $k=0$ along the edge.  

In order to have the edge bands lie within the bulk gap, the right movers and must convert into left movers at the Brillouin zone boundary $k=\pm\frac{\pi}{a}$ where $a$ is the lattice spacing.  This requires that at $k=\pi$ the left and right movers must intersect in groups that transform under the same representation of symmetry:
$\gamma_{L/R,a}\rightarrow \(O_g\)_{a,b}\gamma_{L/R,b}$ where $O_g$ is an orthogonal $N\times N$ representation of the group element $g\in G$.  However, the physically consistent orthogonal representations of $G$ are discrete (even for continuous symmetry groups like $U(1)$) and cannot continuously evolve as a function of edge momentum $k$.  Hence, we may choose a basis at $k=0$ where left and right movers transform identically under symmetry, permitting the mass term: $M\sum_{a=1}^Ni\gamma^T_{L,a}\gamma_{R,a}+\text{h.c.}
$
which trivially gaps the edge - implying that the bulk cannot be an SPT.

\section{Review of 2D Topological Superconductors with $\Z_2$ Symmetry \label{app:2DTSC}}
The three classes of 2D fermion SPT phases (projective, fractional charge, and bosonizable), described in the main text, are all neatly illustrated by a family of 2D topological superconductors (TSCs) with a unitary $\Z_2$ symmetry. These examples are closely related to those introduced in Ref.~\onlinecite{QiZ2TSC}. The $\Z_2$ symmetry can act on, for example, orbital or layer degrees of freedom (while time-reversal symmetry is perhaps a more natural discrete symmetry in electron systems, its antiunitary character unecessarily complicates the theoretical discussion).  The SPT phases of these TSCs are labeled by an integer, $\mathcal{N}$, and can be thought of as $\mathcal{N}$ copies of $p+ip$ superconductors made of spinless (or spin polarized) electrons that are invariant under $\Z_2$, and $\mathcal{N}$ copies of $p-ip$ superconductors made of electrons that obtain a $(-1)$ phase under the $\Z_2$ symmetry (i.e. have an integer unit of $\Z_2$ ``charge").  

The edge of the $\mathcal{N}^\text{th}$ phase contains $\mathcal{N}$ right-moving $\Z_2$-neutral Majorana modes, $\gamma_{R,a}$ ($a=1\dots \mathcal{N}$) and $\mathcal{N}$ left-moving $\Z_2$-charged Majorana modes, $\gamma_{L,a}$.  In non-interacting systems all backscattering terms $i\gamma_{R,a}\gamma_{L,b}$ are symmetry forbidden, and hence the gapless nature of the edge is ensured by symmetry for any integer $\mathcal{N}$.  In interacting systems, however, phases whose $\mathcal{N}$-value differs by a multiple of 8 become topological equivalent - indicating that the integer ($\Z$) classification of interactions collapses to a $\Z_8$ interacting classification.\cite{FidkowskiKitaev,QiZ2TSC}

To see how this works, let us consider the properties of a superconducting vortex ($\pi$-flux) for various $\mathcal{N}$.

\subsection{${\mathcal{N}=1}$, Projective f-SPT}
Consider first, the minimal case $\mathcal{N}=1$.  A superconducting vortex in a single $p+ip$ superconductor carries an unpaired majorana zero-mode, $\gamma$ (and similarly for $p-ip$ superconductors). Then, a vortex in the $\mathcal{N}=1$ TSC carries two zero modes: a $\Z_2$-neutral mode $\gamma_+$ from the $p+ip$ layer and a $\Z_2$-charged mode $\gamma_-$ from the $p-ip$ layer.  This pair of zero modes defines a complex fermion $f = \frac{\gamma_++i\gamma_-}{2}$ and an associated two-level system with states $|\pm \>$ defined by: $i\gamma_+\gamma_-|\pm\> = \pm |\pm\>$, interrelated by $|+\> = f^\dagger |-\>$.  Since $|+\> = f^\dagger |-\>$, the states $|\pm\>$ have opposite fermion parity.  On the other hand, the $\Z_2$ symmetry generator interchanges these states: $\Z_2:|\pm\>\rightarrow |\mp\>$.  To see this, denote  the generator of $\Z_2$ symmetry as $g$, then: $g\(i\gamma_+\gamma_-\)g = i\gamma_+(-\gamma_-)$ (alternatively $gfg = f^\dagger$).  

Moroever, when acting on the vortex zero modes, the $\Z_2$ symmetry anticommutes with the ever-present fermion parity ($\Z_2^F$) ``symmetry" generated by operator $P_F$ ($P_F f P_F = -f$).  Whereas $P_F$ and $g$ ordinarily commute, when acting on the vortex core zero modes of the $\mathcal{N}=1$ phase, these operators \emph{anti-commmute} ($P_FgP_Fg = -1$), since $g$ flips the fermion parity of the states $|\pm\>$.

This projective action of symmetry ensures that no symmetry preserving perturbation can gap out the vortex core zero-modes, and there is a two-fold symmetry protected degeneracy in every vortex core - indicating that the $\mathcal{N}=1$ phase is an example of a projective f-SPT.

\subsection{${\mathcal{N}=2}$, Fractional Symmetry Charge f-SPT }
We now review that the $\mathcal{N}=2$ phase is a fractional symmetry charge f-SPT. Picturing the $\mathcal{N}=2$ phase as a decoupled stack of chiral superconductors would suggest that a vortex contains four Majorana modes, including two $\Z_2$-neutral modes: $\gamma_{+,a=1,2}$ and two $\Z_2$-charged modes: $\gamma_{-,a=1,2}$.  However, the 4-fold degeneracy associated with these Majorana modes can be completely lifted without breaking symmetry, for example by adding the $\Z_2$ symmetric perturbations: $i\gamma_{+,1}\gamma_{+,2}$, and $i\gamma_{-,1}\gamma_{-,2}$.

Despite having no symmetry protected degenerate modes, however, the vortex still has nontrivial symmetry properties.  To see this, a useful mental device is to imagine promoting the $\Z_2^F$ fermion parity to a full $U(1)$ symmetry defined at the edge by:
\begin{align}
\begin{pmatrix}\gamma_{R/L,1} \\ \gamma_{R/L,2} \end{pmatrix} \overset{U(1)}{\longrightarrow} \begin{pmatrix} \cos \theta & \sin \theta \\ -\sin\theta & \cos\theta \end{pmatrix}\begin{pmatrix}\gamma_{R/L,1} \\ \gamma_{R/L,2} \end{pmatrix}
\end{align}
(and similarly in the bulk) and then later break this extra $U(1)$ symmetry back down to $\Z_2^F$.

With the enhanced $U(1)$ symmetry, the $\mathcal{N}=2$ phase is nothing but a bilayer quantum Hall (QH) state consisting of a $\nu=1$ QH state of $\Z_2$-neutral particles and a $\nu=-1$ QH state of $\Z_2$-charged particles.  This analogy, along with simple Laughlin flux-threading arguments, immediately shows that the $\pi$-flux binds a half unit of $\Z_2$ charge, i.e. transforms by a phase of $\pm i$ under the $\Z_2$ symmetry (note the $\pm$ signs can be interchanged by locally binding a $\Z_2$-charged fermion to the $\pi$-flux, but the integer charged fermions cannot screen the residual half-$\Z_2$-charge).

This fractional charge is unchanged upon breaking the enhanced $U(1)$ symmetry back down to the original fermion parity, $\Z_2^F$.  This implies that the vortex in the $\mathcal{N}=2$ phase carries a fractional charge of the $\Z_2$ symmetry  indicating that this phase is an example of a fractional charge f-SPT.

\subsection{$\mathcal{N}=4$, Bosonizable f-SPT}
Viewing the $\mathcal{N}=4$ state as two copies of the $\mathcal{N}=2$ phase immediately shows that symmetry acts non-projectively on the vortex in the $\mathcal{N}=4$ state, and that the vortex carries integer $\Z_2$-charge.  In fact, we can always consider the vortex to be $\Z_2$-neutral by combining it with $\Z_2$ charged fermions to ``screen" its charge (note that this binding of fermions does not change the bosonic statistics of the vortex, due to the mutual statistics between fermions and the vortex).

The trivial symmetry properties of the vortex imply that fermionic degrees of freedom do not play an important role in determining the topological properties of this phase - and that the fermion excitations can be gapped at the edge without breaking symmetry.  However, the phase is not a trivial insulator.  

To see this, imagine inserting a symmetry flux (a $\pi$-flux that is seen only by the $\Z_2$-charged fermions).  As for the $\mathcal{N}=2$ phase, it is again conceptually convenient to enhance the fermion parity to a full $U(1)$ rotation symmetry.  Then, we may view the model as a stack of $\nu=\pm 2$ QH phase of $\Z_2$-charge $\pm 1$ particles.  In particular, the symmetry $\pi$-flux binds integer $\Z_2$ charge through the QH response.  Such a $\pi$-flux bound to unit charge object formally has semionic statistics (i.e. the edge operator corresponding to the symmetry flux will have semionic statistics, or equivalently, if we gauge the symmetry so that the symmetry flux becomes a gapped dynamical quasiparticle excitation, it will be a semion).  As for the $\mathcal{N}=2$ phase, this property of the symmetry flux is not altered when we dispense with the enhanced $U(1)$ symmetry breaking it back down to the original $\Z_2^F$ fermion parity.

This semionic statistics of symmetry-flux is the defining property of the purely bosonic SPT with $\Z_2$ symmetry introduced by Levin and Gu\cite{LevinGu}.  In fact, the $\mathcal{N}=4$ state is topologically identical to the Levin-Gu bosonic SPT.  Hence, we see that the $\mathcal{N}=4$ phase is an example of a bosonizable f-SPT.

\subsection{$\mathcal{N}=8$, Trivial phase}
Using the now familiar tact of viewing the $\mathcal{N}=8$ phase as two copies of $\mathcal{N}=4$ phases, we see that there are no nontrivial properties of either the vortex or $\Z_2$-symmetry flux in this phase.  This implies that the $\mathcal{N}=8$ TSC is topologically trivial in the presence of interactions, and moreover that phases with $\mathcal{N}$ differing by a multiple of eight are topologically equivalent. 

The remaining distinct phases with $\mathcal{N}=3,5,6,7$ are combinations of other $\mathcal{N}$-phases described above, and hence are just combinations of projective and fractional charge f-SPT, and/or bosonic SPTs.


\section{Details of fermionic decorated domain wall construction \label{app:DDW}}
\subsection{SPT Character of DDW model}
Suppose we can construct a Hamiltonian with symmetry group $\Z_2\times G$, for which the DDW wave-function, $|\Psi_\text{DDW}\> = \sum_{\mathcal{C}}|\psi^{(F)}_\text{G-SPT}(\mathcal{C})\>\otimes|\tau^z(\mathcal{C})\>$, is the unique ground-state. Here, $\mathcal{C}$ is a domain wall configuration corresponding to $\tau^z$ wavefunction $|\tau^z(\mathcal{C})\>$, and $|\psi^{(F)}_\text{G-SPT}(\mathcal{C})\>$ is a fermionic wavefunction with fermions along domain walls of $\mathcal{C}$ being in the 1D f-SPT phase protected by symmetry $G$, and fermions away from the domain walls being in a trivial product state.  Then, we can readily see that $|\Psi_\text{DDW}\>$ represents a nontrivial SPT ground state.  First, $|\Psi_\text{DDW}\>$ preserves the $G$ symmetry since $\tau^z$ transform trivially under $G$, and  $|\psi_\text{G-SPT}(\mathcal{C})\>$ is manifestly $G$-invariant for all $\mathcal{C}$. Second, since the $\tau^z$ variables are in a paramagnetic state $|\Psi_\text{DDW}\>$ preserves the $\Z_2$ symmetry.  

Lastly, $|\Psi_\text{DDW}\>$ cannot be deformed into the trivial paramagnetic state $|\Psi_\text{trivial-PM}\>= \sum_{\mathcal{C}}|\psi^{(F)}_\text{triv}(\mathcal{C})\>\otimes|\tau^z(\mathcal{C})\>$ without breaking either $\Z_2$ or $G$, where $|\Psi_\text{triv}(\mathcal{C})\>$ is a trivial product state of the fermionic degrees of freedom.  To see this note that, unless $G$ is broken, upon deforming $|\psi_\text{G-SPT}(\mathcal{C})\>$  into $|\psi^{(F)}_\text{triv}\>$ one necessarily passes through a DW phase transition where low-energy excitations emerge on a DW of length $L_\text{DW}$ with characteristic energy gap $L_\text{DW}$.  Moreover, in order to preserve $\Z_2$ the bulk wavefunction inevitably contains $\tau^z$ configurations with DWs of arbitrary length, up to the system size $L$.  Together this implies that deforming $|\Psi_\text{DDW}\>$ into $|\Psi_\text{trivial-PM}\>$ inevitably creates low-energy excitations excitations with energy $\sim \frac{1}{L}$ that vanishes in the thermodynamic limit, i.e. implies that $|\Psi_\text{DDW}\>$ and $|\Psi_\text{trivial-PM}\>$ are separated by a bulk phase transition.

\subsection{Explicit form of ``decoration" terms}
To explicitly write the decoration terms, it is convenient to represent the complex fermions by Majorana fields $\chi_{r,a} = c_{r,a}+c_{r,a}^\dagger$ and $\eta_{r,a}=-i\(c_{r,a}-c_{r,a}^\dagger\)$, such that $c_{r,a}=(\chi_{r,a}+i\eta_{r,a})/2$.  Eq.~\ref{eq:HDecorate} can be easily written as an explicit sum of local terms:
\begin{align} H_\text{decorate} =\sum_{\<PP'\>,r}&-\frac{(1+\tau^z_P)(1-\tau^z_{P'})}{2} i\chi_{r,a}\eta_{r+d_{PP'},a}~-
\nonumber\\
&-\frac{(1-\tau^z_P)(1+\tau^z_{P'})}{2}i\chi_{r+d_{PP'},a}\eta_{r,a} ~+
\nonumber\\
&+\frac{1+\tau_P^z\tau_{P'}^z}{2}i\chi_{r,a}\eta_{r,a}
\label{eq:HDecorateExplicit}
\end{align}
where $P$ and $P'$ are neighboring plaquettes, such that the line connecting $P$ to $P'$ bisects the bond connecting $r$ and $r+d_{PP'}$.  An arbitrary convention must be chosen to fix the orientation of $d_{PP'}$: for example, we will take the Kitaev $d_{PP'}$ oriented clockwise around plaquettes with $\tau_P^z=-1$.  The last term simply projects sites away from $\tau-$DWs into a trivial empty state.

\subsection{Odd fermion parity of Kitaev ring and related complications \label{app:OddParity}}
The TF terms written above face serious problems for $\mathcal{N}\neq 0,4$.  These issues stem from the fact that the ground-state of the Kitaev chain with periodic boundary conditions has odd fermion parity compared to the ground-state of a trivial chain of the same length.  To see this, note that the total fermion parity of an $L$-site chain is given by:
\begin{align}
P_F &= \prod_{j=1}^L (-1)^{c^\dagger_{j,a}c_{j,a}} = \prod_{j=1}^L i\chi_{j,a}\eta_{j,a} =
 -\prod_{j=1}^L i\chi_{j,a}\eta_{j+1,a}
\end{align}
and recall that the trivial chain ground-state has $i\chi_{j,a}\eta_{j,a}=1 \forall j$, whereas the topological one has $i\chi_{j,a}\eta_{j+1,a}=1 \forall j$. Here site indices $j$ are taken modulo $L$.

Hence, for $\mathcal{N}=1$, the operators: $\Psi_\mathcal{C,P}=|F_{\tilde{\mathcal{C}}(P)}\>\<F_{\mathcal{C}}|$ appearing in Eq.~\ref{eq:HTF} are fermionic whenever the parity of the number of domains domains differs between $\tilde{\mathcal{C}}(P)$ and $\mathcal{C}$.  These operators are not self-hermitian and square to zero: $\Psi_{\mathcal{C},P}^2 = \<F_{\mathcal{C}}|F_{\tilde{\mathcal{C}}(P)}\>\Psi_{\mathcal{C},P}=0$, and anticommute on distant plaquettes: $\{\Psi_{\mathcal{C},P},\Psi_{\mathcal{C},P'}\}=0$ ($P\neq P'$).
Hence, such terms are highly non-local and cannot appear in the Hamiltonian.  This problem can be surmounted by implementing local constraints that prevent the global parity of domain walls from changing, however this necessarily spoils the exact solvability of the model.  Therefore the DDW $\mathcal{N}=1$ construction fails.

For general $\mathcal{N}$, $\Z_2$ domain parity changing terms take the form of $\mathcal{O}_{\mathcal{C},\mathcal{N}}\sim\prod_{a=1}^\mathcal{N} \Psi_{\mathcal{C},P,a}$.  For $\mathcal{N}$ odd, $\mathcal{O}_{\mathcal{C},\mathcal{N}}$ are again fermionic and the construction fails just as for $\mathcal{N}=1$.
For even $\mathcal{N}$, on the other hand, all terms in Eq.~\ref{eq:HTF} are bosonic.  Despite this, the construction still fails for $\mathcal{N}=2,6$.  

To see this, consider the $\mathcal{N}=2$ case.  To ensure that we obtain a nontrivial SPT phase the $G$ symmetry must prevent terms like $i\chi_{r,1}\chi_{r,2}$ and $i\eta_{r,1}\eta_{r,2}$ that can be added to adiabatically deform the Ising domain walls from decorated to trivial without closing a gap.  However, the operators $\mathcal{O}_{\mathcal{C},2}\sim\prod_{a=1,2} \Psi_{\mathcal{C},P,a}$ have the same form, and must therefore also be symmetry forbidden.  Hence, the $\mathcal{N}=2$ DDW construction inherently breaks the $G$ symmetry, and fails to produce a nontrivial SPT phase.  Similar arguments rule out $\mathcal{N}=6$. 

These problems are absent, however, when $\mathcal{N}=4$.  Here, it is possible to ensure that all trivializing bilinear terms (e.g. $i\chi_{r,a}\chi_{r,b}$) are odd under symmetry, without ruling out the four-fermion operators $\mathcal{O}_{\mathcal{C},\mathcal{N}}$ necessary to implement the DDW construction. Consequently, the DDW construction succeeds at providing a commuting projector Hamiltonian with nontrivial SPT character only for $\mathcal{N}=4$. One suitable choice of a symmetry group $G$ that can protect four copies of the Kitaev chain, which we will explore in the next sections, is to take $G=\Z_2^2$, and choose each of the four Kitaev chains to have a different $G$-charge.

\subsection{Symmetric commuting projector character of $\mathcal{N}=4$ model \label{app:N4Model}}
In this Appendix, we explicitly verify the commuting projector structure of the bosonizable $\mathcal{N}=4$ TSC described in the main text. First, it is straightforward to verify that Eqs.~\ref{eq:HDecorate} and \ref{eq:HTF} can by recast in the form of a commuting projector Hamiltonian with a unique ground state.
The terms in Eq.~\ref{eq:HDecorate} for fixed $\{P,P',r\}$, and those in Eq.~\ref{eq:HTF} for fixed $P$ square to unity, and hence have eigenvalues $\pm 1$.  Then, for any such term, $\mathcal{O}$, $\Pi_{\mathcal{O},\pm} =\frac{1}{2}\(1\pm \mathcal{O}\)$ are projection operators having eigenvalues $\{0,1\}$ and satisfying $\Pi_{\mathcal{O},\pm}^2=\Pi_{\mathcal{O},\pm}$. Secondly, it is easy to verify by inspection, that all terms respect the required $\Z_2^3$ symmetry. 

It then remains to check that the projectors commute with one another.  By inspection the decoration terms in Eq.~\ref{eq:HDecorate} commute with each other.  Also, by design, the decoration terms also commute with the terms in Eq.~\ref{eq:HTF}, since the transverse field terms preserve the low energy decorated DW subspace of $H_\text{decorate}$, and vanish when acting on improperly decorated DW configurations.  Lastly, starting with any spin configuration $\mathcal{C}$ and fermion state $|F_\mathcal{C}\>$ and flipping two different plaquettes, $P,P'$, using the terms in Eq.~\ref{eq:HTF}, produces the same spin configuration, and the fermion configurations that agree up to a potential overall phase: $|F_{\mathcal{C}(PP')}\>=e^{i\alpha}|F_{\mathcal{C}(P'P)}\>$.
Since the terms in Eq.~\ref{eq:HTF} have the same action on all four flavors of fermions we may write $e^{i\alpha} = \prod_{a=1\dots 4} e^{i\alpha_a} = \(e^{i\alpha_1}\)^4$.

To compute this phase, note that, since all terms in the Hamiltonian are real, the Hamiltonian satisfies an extra time-reversal-like anti-untiary symmetry that acts like complex conjugation $\mathcal{T}=K$ in the fock basis of the site fermions $c_{r,a}$ (i.e. $\T c_{r,a}\T^{-1} = c_{r,a}$, $\T\chi_{r,a}\T^{-1}=\chi_{r,a}$, $\T\eta_{r,a}\T^{-1}=-\eta_{r,a}$).  This, anti-unitary symmetry requires that $\alpha_a\in {0,\pi}$, implying that $e^{4i\alpha_1} =1$, i.e. that the TF terms all commute.

These considerations verify that the DDW construction for the $\mathcal{N}=4$ TSC protected by $\Z_2^3$ symmetry has all the desired properties for a MB localizable parent Hamiltonian.

\subsection{Equivalence of $\Z_2^3$ fermionic and bosonic phases for $\mathcal{N}=4$ \label{app:InteractingEndStates}}
In this Appendix, we show explicitly that the $\mathcal{N}=4$ TSC phase with $\Z_2^3$ symmetry is equivalent, with interactions, to a bosonic SPT phase. 

Let us write the symmetry group as $\Z_2^A\times\Z_2^B\times\Z_2^C$ to label the different $\Z_2$ subgroups.  As described in the main text, a consistent, solvable DDW model of the $\mathcal{N}=4$ phase is obtained by decorating with four copies of Kitaev chains.  A nontrivial SPT phase is obtained by taking each copy to have a distinct charge under $G=\Z_2^B\times\Z_2^C$: $\{(+,+),(+,-),(-,+),(-,-)\}$.  

Without interactions, cutting open the decorating 1D SPTs reveals four Majorana end states on each end - indicating $2^2=4$-fold degeneracy.  With interactions, we can gap out half of these degrees of freedom by adding the interaction term: $-V(i\gamma_{++}\gamma_{+-})(i\gamma_{-+}\gamma_{--})$ that  favors making the fermion parities $P_2 = (i\gamma_{++}\gamma_{+-})$ and $P_2=(i\gamma_{-+}\gamma_{--})$ the same.  There is still a two-fold degeneracy with this interaction term, spanned by $|\pm\>\equiv |P_1=\pm,P_2=\pm\>$.  $\Z_2^B$ is diagonal in the $|\pm\>$ basis, and we may represent its generator as $g_A|\pm\>=\pm|\pm\>$.  On the other hand, we see that the $\Z_2^B$ generator changes the parity values: $g_BP_{1,2}g_B = -P_{1,2}$.  Since $g_B$ interchanges $|\pm\>$ it necessarily anti-commutes with $g_C$ on the end-state manifold. This projective realization of  the $G$ symmetry protects the remaining two-fold degeneracy. 

The above considerations show, that the $\Z_2^A$ DW ends carry a ``spin-$1/2$" like degree of freedom on which $\Z_2^{B,C}$ act like $\pi$-spin rotations around orthogonal axes (e.g. are represented by the Pauli matrices: $\sigma^{x,y}$ respectively). This is a discrete analog of Haldane's famous gapped spin-1 chain, and is precisely the projective realization of $\Z_2^2$ symmetry obtained by the purely bosonic 1D SPT discussed in Refs.~\onlinecite{ChenDDW,Bahri2013}.

Consequently, we see that, with interactions, this 1D chain is topologically equivalent to the bosonic $\Z_2\times\Z_2$ chain plus a trivial gapped fermion insulator.  Consequently, the corresponding $\Z_2\times G$ DDW construction is topologically equivalent to the bosonic SPT state described in Ref.~\onlinecite{ChenDDW} plus a decoupled trivial fermion insulator.  However, the fermionic and bosonic constructions are qualitatively distinct - since the fermionic DDW approach taken in this paper produces a state with highly entangled bosonic and fermionic degrees of freedom.  We will see at the end of the following section that this qualitative distinction makes the fermionic DDW state closer, in some sense, to the noninteracting topological superconductor than to the purely bosonic one.

\subsection{Edge properties\label{app:DDWEdge}}
In this Appendix, we work out the edge properties of the MB-localizable $\mathcal{N}=4$ TSC phase protected by $\Z_2^3$ symmetry, and verify that they have the expected SPT properties. For clarity, let us distinctly label the three different $\Z_2$ subgroups of the $\Z_2^3$ symmetry as $A$, $B$, and $C$ such that the Ising spins $\tau^z$ transform under $\Z_2^A$, and the fermions $c_{r,a}$ transform under $\Z_2^B\times\Z_2^C$.

To introduce an edge, consider the above $\mathcal{N}=4$ model with spatially uniform couplings in a finite sample with open boundary along the $x$ direction.  In this section we identify possible edge phases of the the ground-state of the spatially uniform $\mathcal{N}=4$ DDW model.  These SPT edge state properties will be inherited by every state in the spectrum for the strongly random MBL version of this model model.  

\subsubsection{Gapped symmetry broken edge states }
One can represent the edge degrees of freedom consists of complex fermions $c_{r,a}$ on the edge sites, and Ising spins on  $\tau^z_{r,r+1}$ on links between the sites on the edge.  At low energies, only the MZM end states of $\tau$ DWs are relevant.  In terms of bulk fermions $c_{r,a}$ the DW zero modes are either $\chi_{r,a}$ or $\eta_{r,a}$ for $\tau^z_{r-1,r}\tau^z_{r,r+1}=\pm 1$ respectively.

A trivially insulating edge can be obtained by placing the boundary spins in an ordered phase with long-range correlations in $\tau^z_{r,r+1}$, at the expense of breaking the $\Z_2$ symmetry at the boundary.  Alternatively, we can produce gapped edge that preserves $\Z_2$ but breaks time-reversal by choosing the edge spins to be eigenstates of $\tau^x_{r,r+1}$, and gapping the low-energy fermionic zero modes with a $\Z_2^T$ symmetry breaking term: $\lambda_B\sum_{a,b}\[i\chi_{r,a}M_{a,b}\chi_{r,b}+i\eta_{r,a}M_{a,b}\eta_{r,b}\]$, where $M_{ab}$ is any suitable mass matrix.

However, an edge that respects the full $\Z_2^3$ symmetry must have $\Z_2^A$ domain walls proliferated, and the residual $\Z_2^B\times \Z_2^C$ symmetry protects some degeneracies of the corresponding DW MZMs.  Hence, the proliferation of DWs containing protected zero modes, required to preserve $\Z_2^A$ symmetry, necessarily produces a gapless edge.

\subsubsection{Fermionic gapless symmetry preserving edge}
Insight into the gapless symmetry preserving edge terminations of the $\mathcal{N}=4$ model can be obtained by considering an antiferromagnetic arrangement of the Ising spins $\tau^z_j = (-1)^j$ where $j$ labels sites along the edge.  This termination breaks $\Z_2^A$, but leaves a residual symmetry of $\tilde{\Z}_2^A=\Z_2^AT_x$ where $T_x$ generates a spatial translation by one lattice spacing along the edge (which we take to be along the x direction).

In the absence of any edge couplings, there are four unpaired Majorana fermions $\gamma_{s,s';j}$ at each site along the edge for this AFM spin arrangement, where $s,s'\in\{\pm 1\}$ label the charges under $\Z_2^B\times\Z_2^C$, such that the edge is highly degenerate.  We may largely lift this degeneracy by adding the symmetry allowed couplings:
\begin{align} 
H_\text{F,edge} &= -\frac{iv}{2}\sum_{j,s,s'}\gamma_{s,s';j+1}\gamma_{s,s';j} 
\nonumber\\
&= v\sum_{s,s'; k>0}\sin k~\gamma_{s,s';k}\gamma_{s,s';-k}
\end{align}
where $\gamma_{s,s';k} = \frac{1}{\sqrt{L}}\sum_{j}e^{ikj}\gamma_{s,s';j}$. Linearizing the dispersion near $k=0,\pi$  edge mode in terms of right (R) and left (L) movers: $\gamma_{ss';R/L,j} = \displaystyle\sum_{k\approx 0/\pi}e^{-i k j}\gamma_{ss';k}$ that transform oppositely under the modified $\tilde{\Z}_2^A$ symmetry.  We can then imagine quantum disordering the AFM edge structure to restore the original $\Z_2^A$ symmetry without gapping out the fermionic edge modes.

These arguments show that the gapless symmetry preserving edge of the $\mathcal{N}=4$ model consists of four copies of nonchiral gapless Majorana fermions, each carrying a distinct combinations of $\Z_2^B\times\Z_2^C$ charges, and verifies that the solvable interacting lattice model based on the DDW construction reproduces the desired edge properties of its free-fermion analog.

\subsubsection{Bosonic Gapless Edge}
As shown in Appendix~\ref{app:InteractingEndStates} above, interactions can reduce the 1D fermionic SPT used to decorate DWs to a bosonic one while preserving the $\Z_2^2$ symmetry.  
To see how this works in terms of the continuum description of the free-fermion edge with four non-chiral gapless Majorana modes, it is useful to use a bosonized description of the gapless fermionic edge.  From the 8 chiral Majorana fermion fields: $\gamma_{s,s',L/R}$, define 4 chiral complex-fermion fields that can be bosonized:
\begin{align}
\psi_{s,L/R} = \frac{1}{2}\(\gamma_{s,+,L/R}+i\gamma_{s,-,L/R}\) \equiv e^{-i\phi_{L/R,s}}
\end{align}
In terms of the boson fields $\phi_{\alpha,s}$ the $\Z_2^3$ symmetry acts as:
\begin{align}
\phi_{R,s}&\overset{\Z_2^A}{\longrightarrow}\phi_{R,s} \nonumber\\
\phi_{L,s}&\overset{\Z_2^A}{\longrightarrow}\phi_{L,s}+\pi \nonumber\\
\phi_{R/L,+}&\overset{\Z_2^B}{\longrightarrow}\phi_{R/L,+} \nonumber\\
\phi_{R/L,-}&\overset{\Z_2^B}{\longrightarrow}\phi_{R/L,-}+\pi \nonumber\\
\phi_{R/L,s}&\overset{\Z_2^C}{\longrightarrow}-\phi_{R/L,s} \nonumber\\
\end{align}

The edge Lagrangian density in the absence of interactions is then:
\begin{align}
\mathcal{L}_\text{edge} = \frac{1}{4\pi}\[\d_x\phi_IK^{IJ}\d_t\phi_J + \d_x\phi_I v^{IJ}\d_x\phi_J\]
\end{align}
where $I,J\in\{(s=+1,R),(-1,R),(+1,L),(-1,L)\}$, $K^{IJ}$ is a diagonal matrix with diagonal entries $\{+1,+1,-1,-1\}$, and $v^{IJ}$ is a non-universal velocity matrix that depends on the edge hopping amplitudes (and is also affected by interactions).

By analogy to Appendix~\ref{app:InteractingEndStates}) above, the edge interaction term $\lambda\int dx \cos\(\sum_{\alpha=L/R,s=\pm 1}\phi_{\alpha,s}(x)\)$ is allowed by symmetry.  When relevant (in the renormalization group, RG, sense), this term gaps out all fermionic excitations at the edge.  However, for weak interactions at the edge (small $\lambda$ and $v^{IJ}$ nearly diagonal) the fermion-gapping term is RG irrelevant - and the fermionic edge described in the previous section is stable to interactions.  

This highlights a qualitative distinction between the fermionic DDW construction here, and the bosonic one of Ref.~\onlinecite{ChenDDW}, though the $\mathcal{N}=4$ topological superconductor produced by the fermionic DDW is topologically equivalent to a purely bosonic SPT accompanied by a spectator trivial insulator of fermions in the presence of interactions.  Since the fermionic DDW approach produces a state with strongly entangled fermionic and bosonic degrees of freedom its symmetry preserving edge state is naturally closer to the free fermion edge with gapless Majorana modes.

\bibliography{MBLSPT}

\end{document}